\documentclass[desactivate]{aa}  

\usepackage{graphicx}
\usepackage{tabularx}
\usepackage{booktabs}
\usepackage{xcolor}
\usepackage{threeparttable}
\usepackage{sidecap}
\usepackage[colorlinks=true,linkcolor=blue,citecolor=blue,urlcolor=blue]{hyperref}
\usepackage{txfonts}
\begin{document}

\title{A new chemical network to explore the disequilibrium chemistry of silicates and iron in hot gas giants}

%\subtitle{I. Overviewing the $\kappa$-mechanism}

\author{Paolo Simonetti\inst{1,2,3}
\and Diego Turrini\inst{2,4}
\and Stavro L. Ivanovski\inst{1,3}
\and Romolo Politi\inst{2,5}
\and Eugenio Schisano\inst{2,5}
\and Sergio Fonte\inst{5}
\and Danae Polychroni\inst{1,2,4}
\and Elenia Pacetti\inst{5}
\and Sergio Molinari\inst{5}
\and Michele Zusi\inst{5}
\and Valeria Cottini\inst{6}
}

\institute{
INAF - Osservatorio Astronomico di Trieste, Via G. B. Tiepolo 11, 34143 Trieste, Italy
\and
National Research Centre for High Performance Computing, Big Data and Quantum Computing (ICSC), Via Magnanelli 2, 40033 Casalecchio di Reno (BO), Italy
\and
Institute for the Fundamental Physics of the Universe (IFPU), Via Beirut 2, 34014 Trieste, Italy
\and
INAF - Osservatorio Astrofisico di Torino, Via Osservatorio 20, 10025 Pino Torinese (TO), Italy
\and
INAF - Istituto di Astrofisica e Planetologia Spaziali, Via del Fosso del Cavaliere 100, 00133 Rome, Italy
\and
Agenzia Spaziale Italiana, Via del Politecnico snc, 00133 Rome, Italy
}

\date{Accepted XXX. Received YYY; in original form ZZZ}

% \abstract{}{}{}{}{} 
% 5 {} token are mandatory
 
\abstract{Chemical equilibrium modeling of gas giant planet atmospheres has shown that the production and condensation of silicates plays an important role in shaping their final composition. However, no chemical kinetics network for these species are available, preventing the opportunity to analyze the impact of chemical disequilibrium processes. In this paper we present \texttt{SILANE}, a new chemical network that include Mg, Si, Fe and 15 carriers of these elements, including MgSiO$_3$ and Mg$_2$SiO$_4$, ready for use in the VULCAN code. We compare the results of our network with a chemical equilibrium model and discuss them in relation to previous results from literature. We then apply our new network to the analysis of a vertically mixed, Solar-composition atmosphere with equilibrium temperatures $T_{eq}$ in the [700, 1600] K range, finding that (i) rock-forming silicates disappears from the upper atmosphere already at 1100 K, (ii) SiH$_4$ reaches concentrations of $10^{-7}$ in the upper ($P\le 1.0$ bar) atmosphere for planets with $T_{eq}\in$ [800, 1100] K and (iii) the amount of oxygen in refractories decrease from 21\% at $T_{eq}$=700 K to essentially zero already at 1000 K, while in the classical equilibrium framework it remains constant up to $T_{eq}$=1300 K. By applying our network to the case of the young hot Jupiter YSES-1 c and comparing both the mixing ratios of volatile species and the enstatite fraction in the silicate mix, we find evidence of a well-mixed atmosphere, possibly with an average $K_{zz} \ge 10^{10}$ cm$^{2}$s$^{-1}$ up to the $10^{-4}$ bar level.}

%By comparing the recent observations of enstatite-rich clouds in the young hot Jupiter YSES-1 c with our results, we interpret their presence as a sign of a weakly mixed atmosphere, since we predict that little to no enstatite should be present in the upper atmosphere at that planet's equilibrium temperature.


\maketitle
%
%________________________________________________________________

\section{Introduction}

Chemical models of exoplanetary atmospheres can be categorized into two large classes: equilibrium \citep[e.g.][]{burrows99,fegley10,visscher10,stock18,woitke18} or kinetics \citep[e.g.][]{moses11,venot12,miguel14,tsai17,drummond20,alrefaie24}. Chemical equilibrium (CE) models search for the set of abundances in a gas mixture corresponding to the absolute minimum of the Gibbs free energy. As input, they require the knowledge of the enthalpy and entropy of the involved chemical species, which are readily available for thousands of species over a large temperature interval, for example in the NASA tables \citep[][]{mcbride02,goos16}. A positive consequence of this modeling choice is that information on the specific reaction pathways are not needed, as only the final state is computed, which greatly reduces the computational cost. An important side effect, however, is that no dynamical, disequilibrium-inducing phenomenon (e.g.~vertical mixing, sedimentation and photochemistry) can be included, since the evolution over time of the gas mixture is not modeled. Chemical kinetics (CK) codes solve this problem by directly integrating over time a set of differential equations representing the concentrations of chemical species in the system until a steady state is reached. This methodology is much more computationally intensive and requires additional information with respect to the CE case, in the form of reaction rates for each specific chemical reaction considered. Since the same molecule is involved in several different reactions, the amount of data required to build a CK model is much larger than that required by a CE code. Experimentally or numerically derived reaction rates are available only for a fraction of the molecules contained e.g.~in the NASA tables, thus chemical CK models for planetary atmospheres are generally limited to few hundred of species at most. Chemical reaction networks for planetary atmospheres generally focus on H, C, N, O and, to a minor degree, S since these are the most cosmically abundant volatile elements. CK models for exoplanets are mostly of this type \citep[e.g.][]{venot12,venot15,drummond16,venot20,mukherjee24,veillet24}. Special ones, tailored to study Solar System gas giants, also include P, Si, Ge and As \citep{fegley94,wang16}, while others expand the treatment of extrasolar hot and ultra-hot Jupiters by including Ti, V \citep{tsai21} and Si \citep[][]{agundez25} volatile carriers. Some of these networks and the underlying data on reaction rates are available in curated repositories such as the NIST Chemical Kinetics Database\footnote{See e.g. \url{https://kinetics.nist.gov/kinetics/}}, UMIST Database for Astrochemistry \citep[][]{millar24} and the Kinetic Database for Astrochemistry \citep[KIDA,][]{wakelam24}. Chemical kinetics modeling is also important when investigating chemistry of AGB stars winds, both in their gas \citep[][]{duari99,cherchneff06,decin18} and in their condensed \citep[][]{gobrecht16,boulangier19,gobrecht22} components. Chemical networks developed to study this type of environment generally include Al, Mg, Si, Ca, Ti and Fe coupled, in the latter case, with detailed descriptions of the nucleation, condensation and surface reactions to trace the bulk growth of dust grains. This modeling framework has been recently adopted also to describe the formation of mineral clouds in hot Jupiter atmospheres \citep{kiefer24}.
%formed by H, C, N, O and S. Work has been done to include some volatile compounds of P, Si, Ge and As \cite{fegley94,wang16}, and Ti and V \cite{tsai21}, but to our knowledge CK models including rock-forming silicates and extensive lists of carriers of other abundant elements such as Mg and Fe are still missing\footnote{See e.g. the Kinetics Database for Astrochemistry \url{https://kida.astrochem-tools.org}}.

Refractory elements, especially the most cosmically abundant Mg, Si and Fe, play an important role in the chemistry of exoplanetary atmospheres by sequestering volatile species into condensates and changing the abundances retrieved from observations \citep[][]{fegley10,lodders10}. Early CE calculations carried out by \citep[][]{burrows99} have shown that, in a Solar-composition atmosphere, around 22\% of all the oxygen is expected to be tied to Mg an Si if the equilibrium temperature $T_{eq}$ of the planet is below $\sim$1200 K. The dominant carriers are expected to be enstatite (MgSiO$_3$) and forsterite (Mg$_2$SiO$_4$) or, if estatite formation is suppressed, forsterite and quartz (SiO$_2$) \citep[][]{visscher10}. Other minerals involving the relatively abundant Ca and Al elements, such as spinel (MgAl$_2$O$_4$), also contribute to deplete the atmospheric oxygen stock. Recently, \cite{fonte23} expanded the investigation of this topic to non-Solar compositions, finding that for moderately more metallic objects ($Z/Z_\odot \sim 8$) with $T_{eq} \le 900$ K the percentage of sequestered oxygen -mainly in Mg(OH)$_2$ and Fe(OH)$_2$- can reach $\sim$40\%. \textcolor{black}{This oxygen deficit impacts the mixing ratio of the observable carriers such as H$_2$O and CO, and, if not taken into account, leads to an erroneous estimate of the elemental ratio for a planetary atmosphere.} Chief among them is the C/O ratio, that is used in conjunction with models of the position of the snow line in the protoplanetary disk to reconstruct the evolutionary history of Jupiter-like exoplanets \citep[][]{oberg11,mordasini16,turrini21,pacetti22,pacetti25}.

The presence of silicate clouds in planetary atmospheres has been confirmed observationally in \textcolor{black}{the hot Jupiter WASP-17 b \citep[][]{grant23,dyrek24} and} the young long-period Jupiters YSES-1 b and c \citep[][]{hoch25}, while \textcolor{black}{iron condensation on the nightside has been invoked to explain compositional differences between the morning and evening terminator of WASP-76 b \citep[][]{ehrenreich20,kesseli21}}. Iron clouds may also explain the low albedo exhibited by HD 209458 b and WASP-43 b \citep{chubb24}. Within our Solar System, the sulfur abundance in the Jupiter's atmosphere seems to indicate the presence of a sizeable stock of refractories, which are expected to play an important role in the sequestration of oxygen \citep[][]{fonte23}. So far, only CE models have been used to investigate these cases, which leaves unsolved degeneracies clearly linked to atmospheric dynamics and outside the reach of equilibrium chemistry. An example is the degeneracy between the grain size and the position of the cloud base \citep[][]{hoch25}. Another issue very likely related to disequilibrium chemistry is the observed depletion of methane \citep[][]{stevenson10,kreidberg18,dyrek24}. While this might be unrelated to refractories condensation, the currently used HCNOS chemical network fails to provide a widely accepted solution. Rock vapor atmospheres, which are expected \citep[][]{miguel11,ito15} and possibly have been observed \citep[][]{zieba22} around hot Super Earths, will also require a new generation of CK models that include the treatment of non-volatile elements. 

In this work, we present a new chemical network for CK models that include 15 oxides and hydrides of the three most cosmically abundant refractory elements, Mg, Si and Fe. \textcolor{black}{We called this network \texttt{SILANE} (SILicates Atmospheric NEtwork).} We carry out our calculations using the VULCAN \citep[][]{tsai17,tsai21} code and we delineate a procedure to treat silicate condensation in a simplified but readily portable and self-consistent way within the framework of this software (Section \ref{sec:model}). We compare this network against CE calculations, performed with FastChem \citep[][]{stock18,kitzmann24} in the [700, 1600] K temperature range and [10$^{-6}$, 10$^{3}$] bar range (Section \ref{sec:comparison}). We then apply this network to the case of a Solar-composition Jupiter-like planet, in order to investigate how Mg, Si and Fe are partitioned among their products at different planetary equilibrium temperatures, and to the recently observed atmosphere of YSES-1 c \citep[][]{hoch25} (Section \ref{sec:results}). In Section \ref{sec:discussion} we discuss our results with respect to both other modeling efforts and recent observations. Finally, in Section \ref{sec:conclusions} we discuss our results and draw some conclusions.

%This paper is structured as follows: in Section \ref{sec:model} we present our new chemical network, in Section \ref{sec:comparison} we compare CK and CE calculations and in Section \ref{sec:results} we apply our network to an hypothetical Solar-composition hot Jupiter and the young long-period exoplanet YSES-1c. In Section \ref{sec:discussion} we discuss our results with respect to both other modeling efforts and recent observations. Finally in Section \ref{sec:conclusions} we discuss our results and draw some conclusions.

%%%%%%%%%%%%%%%%%%%%%%%%%%%%%%%%%%%%%%%%%%%%%%%%%%%%%%%%%%%%%%%%%%
%%%%%%%%%%%%%%%%%%%%%%%%%%%%%%%%%%%%%%%%%%%%%%%%%%%%%%%%%%%%%%%%%%

\section{Model}
\label{sec:model}

\subsection{The chemical kinetics framework}
\label{ss:vulcan}

The chemical network and the procedure to obtain a self-consistent condensation of silicates have been devised to work with the CK open source code VULCAN \citep[][]{tsai17,tsai21}. While we direct the reader to the reference papers for the details, we recall here the general modeling framework adopted for clarity.

In general, for the chemical reaction
\begin{equation}
iA + jB \rightarrow pC + qD,
\end{equation}
where $i$, $j$, $p$ and $q$ are the stoichiometric coefficients, the reaction rate $v$ is:
\begin{equation}
\label{eq:total_reaction_rate}
v = k_f(T) N_A^i N_B^j - k_r(T) N_C^p N_D^q
%[A]^i[B]^j - k_r(T) [C]^p[D]^q,
\end{equation}
where $k_f$ and $k_r$ are, respectively, the temperature-dependent forward and reverse reaction rate coefficients, and the square brackets indicate the number density of the involved species. $v$ is directly related to the change over time of the molar concentration of the involved species, being equal to the time derivative of \textcolor{black}{the number density of each species} $N_X$, $X \in \{A, B, C, D\}$, divided by the stoichiometric coefficient. The dependence of $k_f$ on temperature is usually expressed using the generalized Arrhenius equation \citep{heng16}:
\begin{equation}
\label{eq:forward_rate}
k_f (T) = \mathcal{A} \biggl(\frac{T}{T_0}\biggl)^b \exp\biggl(-\frac{E_a}{T}\biggl),
\end{equation}
where $\mathcal{A}$ (in units of s$^{-1}$, cm$^3$s$^{-1}$ and cm$^6$s$^{-1}$ respectively for the first, second and third order reactions), $b$ and $E_a$ (which is the activation energy divided by the specific gas constant) are derived from experimental data or theoretical calculations and tabulated in repositories such as the NIST Chemical Kinetics Database\footnote{\url{https://kinetics.nist.gov/}} or the Kinetics Database for Astrochemistry\footnote{\url{https://kida.astrochem-tools.org/}}. $T_{0}$ is a standard-state temperature, here equal to 298 K. The reverse coefficient $k_r$ is instead calculated using the relation:
\begin{equation}
\label{eq:reverse_rate}
\frac{k_f}{k_r} = K_{\text{eq}} \biggl(\frac{k_BT}{P_0}\biggl)^{\Delta\mu},
\end{equation}
where $K_{\text{eq}}$ is the dimensional equilibrium constant, $k_B$ is the Boltzmann constant, $P_0$ is the standard-state pressure (equal to 1 bar) and $\Delta\mu = i + j - p - q$ (i.e. the difference between the stoichiometric coefficients of the left- and right-hand side of the chemical reaction). The equilibrium constant is equal to:
\begin{equation}
K_{\text{eq}} = \exp\biggl(\frac{-\Delta G^0}{RT}\biggl),
\end{equation}
where $\Delta G^0 = \Delta H^0 - T \Delta S^0$ is the change in the standard Gibbs free energy when reactants are transformed into products and $R$ is the universal gas constant. $\Delta H^0$ and $\Delta S^0$ are the changes in the standard enthalpy and standard entropy and are calculated using the NASA9 polynomial coefficients \citep{mcbride02, goos16}, which are valid in the 200-6000 K interval.

By considering all the reactions involving a certain species $X$ it is possible to define a production $\mathcal{P}_X$ and a loss $\mathcal{L}_X$ rate from the right-hand side terms of Eq.~\ref{eq:total_reaction_rate}. $\mathcal{P}_X$ and $\mathcal{L}_X$ can then be used to track the concentration of $X$ in the gas mixture via the mass continuity equation
\begin{equation}
\label{eq:vulcan}
\frac{\partial N_X}{\partial t} = \mathcal{P}_X - \mathcal{L}_X - \frac{\partial \phi_X}{\partial z},
\end{equation}
where transport term $\phi_X$ along the vertical direction $z$ is included. 
%Vertical mixing has the form
In VULCAN, the transport term is the sum of advective, eddy, molecular and thermal diffusion, and is written in the form
\begin{multline}
\label{eq:transport_term}
\phi_X = N_X v_z - K_{\text{zz}} N_{\text{tot}} \frac{\partial (N_X/N_{\text{tot}})}{\partial z} +\\- D_X  \biggl( \frac{\partial N_X}{\partial z} + N_X \biggl( \frac{1}{H_X} + \frac{1+\alpha_T}{T} \frac{\partial T}{\partial z}\biggl) \biggl),
\end{multline}
where $v_z$ is the vertical wind speed, $K_{\text{zz}}$ and $D_X$ are the eddy and molecular diffusion coefficients, respectively, \textcolor{black}{$N_{\text{tot}}$ is the total number density of gas evaluated using the ideal gas law,} $H_X$ is the molecular scale height and $\alpha_T$ is the thermal diffusion factor. In this work we will keep the advection velocity set to 0, while we will test different values for $K_{zz}$. \textcolor{black}{$D_X$ is calculated for each species as
\begin{equation}
\label{eq:molecdiff}
D_X=D_{CH4}\sqrt{(m_{CH4}/m_X)(m_{H2}+m_X)/(m_{H2}+m_{CH4})},
\end{equation}
%$D_X=D_{CH4}\sqrt{m_{CH4}/m_X(m_{H2}+m_X)/(m_{H2}+m_{CH4})}$,
with $D_{CH4}=2.2965\times10^{17}T^{0.765}/N$ \citep[][]{marrero72}. $\alpha_T$ is equal to -0.1, 0.145 and 0.25 for H, He and all the other species, respectively.}

Once formed, certain species are allowed to condense. In order to decide whether this happens, the partial pressure of each condensable species is compared with the saturated vapor pressure at the temperature of each grid point and the related condensation rate is calculated using the mass balance equation \citep[][]{seinfeld16}
\begin{equation}
\label{eq:condensation}
\frac{dN_X}{dt} = -\frac{D_X m_X}{\rho_p r^2_p}(N_X-N_X^{sat})N_X^{sat}
\end{equation}
where $D_X$, $m_X$ and $N_X^{sat}$ are the molecular diffusion, mass and saturation concentration of the species $X$, and $\rho_p$ and $r_p$ are the density and radius of the condensate particle. If the condensate $X$ is present, but its saturation vapor pressure is higher than its gas phase concentration, evaporation will occur with a rate calculated using the same Eq.~\ref{eq:condensation}.

The VULCAN code operates in two stages. First, CE abundances are calculated by using a modified version of the FastChem \citep[][]{stock18} code \textcolor{black}{(hereafter called FastChem-Vulcan)}, that uses the full NASA9 polynomial coefficients rather than the reduced 5-terms polynomial of the standalone FastChem distribution. Second, these CE abundances are processed by the CK core of the VULCAN model. In essence, a set of equations in the form of Eq.~\ref{eq:vulcan}, one for each chemical species at each layer of the vertical grid representing the atmospheric column, is integrated by means of the Rosenbrock method until a steady state is reached. Aside for the vertical mixing and the condensation of certain chemical species, VULCAN can also take into account the effects of the photo- and ion-chemistry and of fluxes at the boundaries of the atmospheric column. These further phenomena, however, are not \textcolor{black}{included in our runs and will not be} discussed further in this work.

\subsection{\textcolor{black}{Grain settling}}
\label{ss:settling}

Condensed particles can settle gravitationally by falling at a given terminal velocity $v_s$. \textcolor{black}{In the published version of VULCAN, the settling speed is always calculated from the Stokes' law, which represents the case of a drag force by a laminar viscous medium \citep[][]{stokes1851}. However, micron-sized particles in the observable layers of Hot Jupiters can also be in the free molecular flow \citep[][]{epstein1924} regime \citep[see e.g.][]{woitke03}. In order to capture both regimes, we modified VULCAN by adopting the interpolated formulation included in ExoLyn \citep[][]{huang24}:
\begin{equation}
\label{eq:falling}
v_s = \frac{\rho_p r_p g}{\rho_g v_{th}} \sqrt{1 + \biggl(\frac{4 r_p}{9 l}\biggl)^2}
\end{equation}
where $\rho_g$ is the background gas density, $v_{th}$ is its thermal velocity and $l$ is the mean free path.}  
%calculated from the Stoke's law:
%\begin{equation}
%\label{eq:falling}
%v_s = \frac{2\rho_p r^2_p g}{9\mu}
%\end{equation}
%where $\mu$ is the atmospheric dynamic viscosity, that in VULCAN are taken from \cite{cloutman00}. \textcolor{black}{VULCAN does not support a condensed particle size distribution in the current release, and in this work we assume a grain size equal to 1 $\mu$m for all condensates.}
%This process removes the condensates from the layer within which condensation occurs, moving them to deeper regions of the atmosphere.

\textcolor{black}{We consider two types of grains in this work: silicate grains with $\rho_p=3.2$ g cm$^{-3}$ and iron grains with  $\rho_p=7.87$ g cm$^{-3}$. As for $r_p$, VULCAN allow to specify a single, constant value for each condensate. In Sect.~\ref{ss:hotjup}, we use $r_p=1$ $\mu$m for both condensates, which represents the order of magnitude of the maximum sizes of magnesium silicate grains in the outer regions of the atmospheres of Brown Dwarfs, as found by both models \citep[e.g.][]{woitke03} and observations \citep[][]{suarez23}. This choice also agrees with models of Hot Jupiter atmospheres \citep[e.g.][]{ormel19}. On the other hand, recent observations of Hot Jupiter atmospheres point to smaller grains, between 0.1 $\mu$m for forsterite and enstatite \citep[][]{hoch25,chubb26} to 0.01 $\mu$m for quartz \citep[][]{grant23,inglis24}. In the conditions treated here, $v_s$ increases monotonically with $r_p$: larger grains settle faster, sequestering more refractories deep into the atmosphere and moving the column further away from the equilibrium chemistry results. Therefore, choosing $r_p=1$ $\mu$m can be considered conservative. In Sect.~\ref{ss:yses1c}, we instead adopt $r_p = 0.1$ $\mu$m, as found by \citet{hoch25}.}

%This parameterization of $v_s$, which is valid in the typical conditions of hot gas giants, is monotone in $r_p$. As such, larger grains settle faster, moving the column further away from the equilibrium chemistry results. In this sense, our $r_p$ choice in Sect.~\ref{ss:hotjup} is conservative with respect to observational data. In Sect.~\ref{ss:yses1c}, we instead adopt $r_p = 0.1$ $\mu$m, as found by \citet{hoch25}.
%This is, of course, a simplification: actual grain sizes likely vary by 4-5 orders of magnitude along the column \citep[][]{arfaux24}, potentially reaching $\sim$100 $\um$m at $P>100$ bar. Thus, our results should be read with this limitation in mind.
%Since, we consider $\mu$m - sized particles or larger grains with higher settling velocities, moving the column further away from the equilibrium chemistry results.

%In practice, for micron-sized grains, the dynamics of condensate is dominated by the turbulent gas motion throughout most of the planetary atmosphere.
Condensate particles are also subjected to the eddy diffusion term in Eq.~\ref{eq:transport_term}, which means that sufficiently small particles can remain suspended in the atmosphere. This framework is similar to that presented in \citet{ackerman01}. For a \textcolor{black}{1 (0.1) $\mu$m} silicate grain ($\rho=3.2$ g cm$^{-3}$) \textcolor{black}{in a 1100 K gas and with a gravitational acceleration equal to that of Jupiter\footnote{$g=2479$ cm s$^-2$}, the falling velocity as calculated from Eq.~\ref{eq:falling} is below the characteristic speed associated with an eddy diffusion of $10^{8}$ cm$^2$ s$^{-1}$ for $P \gtrsim 0.01$ (0.001) bar.} In practice, under the assumptions made to study the cases presented in this paper, we expect to find both silicates and metallic iron suspended throughout \textcolor{black}{most} the column, rather than settled at the bottom.
%\textcolor{black}{Smaller grains can stay aloft at even smaller pressures: a 0.1 $\mu$m particle would not settle even at $P \sim 0.001$ bar.} 
%For micron-sized silicate grains ($\rho=3.2$ g cm$^{-3}$), the falling velocity as calculated from Eq.~\ref{eq:falling} is $2.5\times10^{-2}$ cm s$^{-1}$, which is two orders of magnitude less than the characteristic speed associated with an eddy diffusion of $10^{8}$ cm$^2$ s$^{-1}$ ($v_{Kzz} \approx K_{zz}/H \simeq 3$ cm s$^{-1}$). The same is true for metallic iron grains ($\rho=7.87$ g cm$^{-3}$), although their $v_s$ is around 2.5 times higher. In practice, under the assumptions made to study the cases presented in this paper, we expect to find both silicates and metallic iron suspended throughout the column, rather than settled at the bottom. This is largely in accord with observations of quartz \citep{grant23}, enstatite and forsterite \citep[e.g.][]{hoch25} and the indirect evidence for Fe condensation \citep[][]{ehrenreich20} in hot gas giants.

\subsection{The new chemical network}
\label{ss:network}

We compiled \texttt{SILANE}, a new chemical network that includes Mg, Si, Fe and the main molecules formed when combining them with O, H and S, by using the available experimental data on the NIST database, taking care of selecting reactions whose rates were evaluated within or near the typical temperature range of hot Jupiter atmospheres. The main goal of this network is to study the partition of oxygen among volatile and refractory carriers within the framework of the VULCAN code. The chemical species that we included reflect this goal, since they are important carriers of both the selected refractories and oxygen \citep[][]{visscher10,fonte23}. \textcolor{black}{As a further check, we re-investigated the same range of pressures and temperatures using FastChem3 \citep[][]{kitzmann24}, confirming previous findings. The results of this analysis are discussed in Appendix A.} \texttt{SILANE} \textcolor{black}{includes} a total of 18 species: Mg, Si, Fe, MgO, MgOH, Mg(OH)$_2$, FeO, FeO$_2$, Fe(OH)$_2$, SiS, SiO, SiO$_2$, SiH, SiH$_2$, SiH$_3$, SiH$_4$, MgSiO$_3$ and Mg$_2$SiO$_4$. \textcolor{black}{The thermodynamical properties of these species are defined in VULCAN by making use of the available NASA9 data, as for the other molecules S-N-C-H-O molecules included in the network.} Fe, MgSiO$_3$ and Mg$_2$SiO$_4$ are allowed to condense, thus exist both as gases and as condensates, while the others exist only in the gas phase. Different phases are accounted for separately. In reality, gas-phase MgSiO$_3$ and Mg$_2$SiO$_4$ do not exist, since they dissociate during evaporation, but we included short-lived virtual gas phases \textcolor{black}{(with the same thermodynamical properties of the real condensed phases)} for these two species in order to make \texttt{SILANE} readily compatible with VULCAN.
%\textcolor{black}{Using virtual silicate gas phases is a framework \citet[][]{ormel19} adopted a similar framework in their condensation modeling, even though they did not coupled it with chemistry.}

The \textcolor{black}{46} reactions linking together these species are reported in the Appendix B, organized by the reaction order: second order reactions in Table \ref{tab:bimolecular}, first order dissociation reactions in Table \ref{tab:onemolecular} and third order reactions in Table \ref{tab:termolecular}. These tables report the coefficients for Eq.~\ref{eq:forward_rate}, together with the temperature interval within which each set of these coefficients has been derived and the related bibliographic reference. We included also reactions whose experimental temperature coverage is sparse, as the available data represent the best understanding of their behavior. In this paper, we use \texttt{SILANE} as an expansion of the \texttt{SNCHO\_photo\_network\_2024} distributed together with the VULCAN code, which already included $\sim$90 species of H, C, N, O and S.
%While the temperature coverage of these reactions in the available databases is sparse, we decided to include them nonetheless as this was the best available option.

\subsubsection{Rock-forming silicates}
\label{sss:silicates}

VULCAN chemical reactions take place exclusively in the gas phase, and phase changes only depend upon the ratio between the partial pressure of the species $X$ and its vapor pressure at the specified temperature. \textcolor{black}{As such, the formation of rock-forming silicates, which involves the formation of a new compound directly in the condensed phase on the surface of condensation nuclei \citep[e.g.][]{helling13}, can be treated only indirectly.} Condensation nuclei are supposed to be always present, i.e. they are not formed kinematically within the model, and the grains of condensates have all the same radius in the sedimentation calculation, which is specified as input. In order to include the dominant rock-forming silicates enstatite and forsterite within this simplified framework, we proceed as follows. \textcolor{black}{First, we substituted FastChem-Vulcan with FastChem3, in order to be able to calculate the amount of condensates in chemical equilibrium. To ensure element conservation, we included in FastChem3 only the species that are part of the chemical kinetics network we are using. Then, we pass the abundances obtained in CE to VULCAN, where condensates are allowed}
%First, we let MgSiO$_3$ Mg$_2$SiO$_4$ to form in thermochemical equilibrium, i.e. during the initialization of VULCAN, \textcolor{black}{as if they were gases}. \textcolor{black}{This is done by including the NASA9 polynomial coefficients for these two species in the list of species treated by FastChem-Vulcan. This methodology does not produce the real CE mixing ratio of enstatite and forsterite, since no consideration of the supersaturation ratio is performed. In other words, enstatite and forsterite are produced by taking into account the minimization of the Gibbs free energy, but not the local pressure and temperature conditions, which are instead evaluated in VULCAN. , we checked if the resulting mixing ratios and temperature dependencies are compatible with those found by other published CE frameworks. Results are reported in Appendix C. The MgSiO$_3$ and Mg$_2$SiO$_4$ thus formed is treated as the initial stock of condensed enstatite and forsterite.}
%Then, we allow these condensates
to evaporate into a virtual gas phase and to rapidly dissociate following the reactions, reported in Table \ref{tab:bimolecular},
\begin{equation*}
\text{(1117)}\qquad \text{Mg}_2\text{SiO}_4 + \text{M} \rightarrow \text{MgO} + \text{Mg}\text{SiO}_3 + \text{M}
\end{equation*}
and
\begin{equation*}
\text{(1119)}\qquad \text{Mg}\text{SiO}_3 + \text{M} \rightarrow \text{MgO} + \text{SiO}_2 + \text{M},
\end{equation*}
where M is the background gas. These reactions have been adapted from the results of \citet[][]{laziri24}, which showed that their reverses take place in sintered mixtures of MgO and SiO$_2$ nanopowders. To check if evaporation conditions are met, the local temperature-pressure $(T,P)$ conditions in the $i$th simulated atmospheric layer are compared with the condensation curves for enstatite
\begin{equation}
\label{eq:enstatite}
%P_{En} = 2.9119 \times 10^{(18.886-\frac{10000}{0.35 T})}
P_{En} = 10^{(13.496-\frac{10000}{0.35\, T}-1.973\,[Fe/H])}
%P_{En} = 10^{(3.279\times10^{-5}\,(T-147.6)^2-5.172\times10^{-2}\,(T-147.6)+3.976)} 
\end{equation}
and forsterite
\begin{equation}
\label{eq:forsterite}
%P_{For} = 2.9119 \times 10^{(16.919-\frac{10000}{0.37 T})},
P_{For} = 10^{(11.238-\frac{10000}{0.37\, T}-2.0\,[Fe/H])},
\end{equation}
as taken from \citet{visscher10}. \textcolor{black}{Pressures are expressed in bar and $[Fe/H]=\log_{10}([Fe]/[H])-\log_{10}([Fe]_\odot/[H]_\odot)$ traces the metallicity. Eqs.~\ref{eq:enstatite} and \ref{eq:forsterite} are valid in the [800, 2500] K range and up to $[Fe/H]=+0.5$. We remark that the exponent term includes a scaling factor, which is needed since the pressure in \citet{visscher10} is the total atmospheric pressure, while in VULCAN is the partial pressure of the condensing gases. This scaling factor has been calculated by supposing that the mixing ratio of the least abundant precursor to form each of the two minerals is comparable to the Si element abundance for enstatite, and half of the Mg element abundance for forsterite. The element abundances used to evaluate the scaling factor are those by \cite{lodders03}, in order to be consistent with \citet{visscher10}.}
%as taken from \textcolor{black}{CE calculations by} \cite[][]{visscher10} \textcolor{black}{by setting the [Fe/H] = 0 and expressing $P$ as a function of $T$. The temperature range of validity of Eqs.~\ref{eq:enstatite} and \ref{eq:forsterite} are not explicitly given. As such, we separately tested them against FastChem3, confirming that they can be applied to gas giant planets with $T_{eq}\in [700, 1600]$ K and $Z=1.0\,Z_\odot$. The comparison is carried on in Appendix C.}

As for the pre-exponential factor $\mathcal{A}$ of the reactions 1117 and 1119, \textcolor{black}{we choose it to be of the same order of magnitude of the physical upper limit given by the kinetic theory of gases for the temperature interval of consideration, $10^{-10}$ cm$^3$ molecule s$^{-1}$. This guarantees a very short chemical timescale ($\sim$$\mu$s) for the dissociation of the virtual gas phase. We also tested two more values for $\mathcal{A}$ ($10^{-11}$ and $10^{-12}$ cm$^3$ molecule s$^{-1}$), finding only negligible differences in the final partition of Mg, Si and Fe in their carrier species.}

%As for the pre-exponential factor $\mathcal{A}$ of the reactions 1117 and 1119, we tested different possible values ($10^{-10}$, $10^{-11}$ and $10^{-12}$ cm$^3$ molecule s$^{-1}$). However, we found only negligible differences in the final partition of the refractories in their carrier species, both in presence and in absence of vertical mixing. We, thus, decided to set the pre-exponential coefficient to $10^{-10}$ cm$^3$ molecule s$^{-1}$. \textcolor{black}{This value is of the same order of magnitude of the physical upper limit for the pre-exponential factor in the collision theory for the temperature interval under consideration, guaranteeing a very short chemical timescale for the dissociation process ($\sim$$10^{-6}$ s).}

One issue we encountered while testing \texttt{SILANE} was that allowing these reactions to proceed in both directions (i.e., to both form and dissociate MgSiO$_3$ and Mg$_2$SiO$_4$) resulted in the accumulation of large amounts of virtual gaseous enstatite and forsterite. This was particularly pronounced at higher temperatures, where CE calculations predict these species should be fully dissociated. This outcome occurred regardless of the chosen $\mathcal{A}$ value. Therefore, we opted to treat these reactions strictly as dissociation processes to avoid \textcolor{black}{unphysical behaviour}. \textcolor{black}{In other words, the reverse of the reactions 1117 and 1119 are not included in \texttt{SILANE}.}

Treating the formation of MgSiO$_3$ and Mg$_2$SiO$_4$ as equilibrium processes is justified in light of the growth speed of dust grains in detailed numerical models of silicate cloud formation in hot gas giants. For example, \cite{kiefer24} reported that the bulk growth of forsterite on TiO$_2$ condensation nuclei ends within 1-10 seconds in the 900-1800 K temperature interval and for pressures in the 0.002-2 bar range. This timescale is much shorter than the vertical dynamical timescale $\tau_{dyn}$ of the atmosphere. The latter can be estimated, for a given vertical mixing strength, as $\tau_{dyn} \approx L^2 / K_{zz}$, where $L=0.316H$\footnote{Generally, $L$ is assumed to be related to $H$. Detailed calculations from \cite{visscher11} showed that it is comprised between $0.1H$ and $0.7H$, depending on the species chosen as tracer. Here we chose an intermediate value within this range, so as that $L^2=10H^2$.} and $H$ is the atmospheric scale height, and for the vertical mixing intensity adopted in this work ($k_{zz}=10^8$ cm$^2$s$^{-1}$), it is comprised between $8.4\times 10^{5}$ s when $T = 1600$ K and $1.6\times 10^{5}$ s when $T = 700$ K. The longest timescales reported by \cite{kiefer24}, which concerns the disappearance of fayalite (FeSiO$_4$) and the transfer of its iron content to Fe$_2$O$_3$ (which are not included in \texttt{SILANE}), are still of the order of a few 10$^5$ s.

\subsubsection{Iron}
\label{sss:iron}

Iron is another abundant refractory element that can potentially sequester oxygen in exoplanetary atmospheres. CE calculations without condensation carried out in \citep[][]{fonte23} showed that most Fe forms Fe(OH)$_2$ when $T_{eq}$ is lower than 1100 K, locking away up to 10-13\% of the oxygen stock at 700 K. On the other hand, previous investigations including condensation concluded that most Fe should be locked away in a condensed elementary form, thus unable to capture other volatile species \citep[][]{visscher10}.
%To decide which of these conflicting results is the right one requires an analysis of the timescales of the two dominant processes -iron(II) hydroxide formation vs pure iron condensation- and the inclusion of atmospheric dynamics, which can impact the steady state solution by moving Fe-rich gas in regions where Fe(OH)$_2$ is favored.
In \texttt{SILANE}, we include both the formation of Fe(OH)$_2$ (reaction 1079 in Table \ref{tab:termolecular}) and its precursor, FeO (reactions from 875 to 883 in Table \ref{tab:bimolecular}), and pure Fe condensation following the relation \textcolor{black}{given in Table 1 of \citet{visscher10}, where $P_{Fe}$ is expressed in bars:}
%derived from \citep[][]{visscher10}
\begin{equation}
\label{eq:Fe_condensation}
%P_{Fe} = 2.9119 \times 10^{(12.33-\frac{10000}{0.48 T})}.
P_{Fe} = 10^{(7.23-\frac{20995}{T})}.
\end{equation}
\textcolor{black}{This expression is valid in the [800, 2500] K range.}

\subsection{Model planet used as a test case}
\label{ss:model_planet}

\begin{figure}
%\isPreprints{\centering}{} % Only used for preprints
\includegraphics[width=0.45\textwidth]{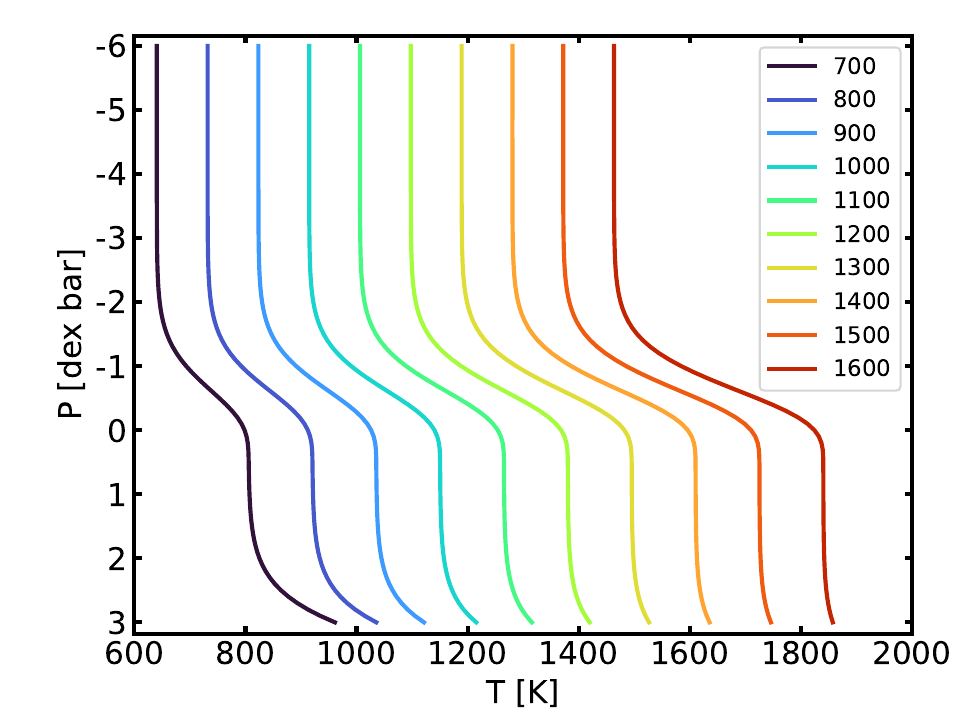}
\caption{The pressure-temperature profiles adopted in this work. The legend reports the $T_{eq}$ of each profile. \label{fig:pt_profiles}}
\end{figure} 

We test our network on a model gas giant exoplanet with mass and radius equal to those of Jupiter, a Solar-composition atmosphere \citep[][Table B.1]{asplund21} and equilibrium temperatures $T_{eq}$ within the [700, 1600] K range. The vertical pressure-temperature profiles adopted are the globally-averaged ones of \cite[][]{guillot10}, derived by choosing an internal temperature of 100 K, an infrared opacity of 0.01 cm$^2$ g$^{-1}$ and a visible opacity of 0.004 cm$^2$ g$^{-1}$, which are adequate to describe the radiative transfer properties of the chosen Solar-composition atmosphere. They are reported in Figure \ref{fig:pt_profiles}. The atmospheric pressure grid is comprised of 90 layers, ten per order of magnitude in the [$10^{-6}$, $10^{3}$] bar interval. The $K_{zz}$ coefficient is taken constant at all altitudes. While real atmospheres usually show vertical variations of the mixing intensity \citep[e.g.][]{massie81} and some General Circulation Model-derived $K_{zz}$ are available for exoplanets \citep[e.g.][]{parmentier13}, we still favored this simpler approach due to the intrinsic uncertainties related to this simplified treatment of the mixing phenomenon. The same model planet is used both during the comparison with the equilibrium chemistry results (Sect.~\ref{sec:comparison}) and when investigating the effects of vertical mixing (Sect.~\ref{sec:results}).

%%%%%%%%%%%%%%%%%%%%%%%%%%%%%%%%%%%%%%%%%%%%%%%%%%%%%%%%%%%%%%%%%%
%%%%%%%%%%%%%%%%%%%%%%%%%%%%%%%%%%%%%%%%%%%%%%%%%%%%%%%%%%%%%%%%%%

\section{Comparison with equilibrium chemistry}
\label{sec:comparison}

First of all, we study the partition of Mg, Si, Fe and O among their carriers as obtained by \texttt{SILANE} when disequilibrium mechanisms are turned off, i.e. by setting to zero the vertical transport term in Eq.~\ref{eq:vulcan} and the gravitational settling of condensed particles. As such, the mixing ratios of all species within each atmospheric layer are functions of the local production and destruction rates only. These mixing ratios can thus be compared directly with those derived by an equilibrium chemistry code, which again depends only on the local thermochemical state of the atmosphere. \textcolor{black}{This test is needed since the network includes non reversible reactions, thus in principle VULCAN is not expected to converge to the equilibrium chemistry solution \citep[][]{vanderschaft13}. As such, deviations to the predicted equilibrium VMR must be quantified in the temperature interval of application. Moreover, we check that no significant accumulation of Mg and Si happens in the virtual gas phase under the standard numerical convergence conditions suggested for VULCAN.}
%\textcolor{black}{The reason for this test is to show that the network is sufficiently complete to produce a reasonable partition of the added elements into their carrier species, i.e. that it is sufficiently extended to ensure a final partition of tracked elements that is compatible with the one found by applying thermodynamical arguments. This need arises from the fact that, in a chemical kinetics code, the allowed reaction pathways are specified in advance: a network that is too small might miss one or more key reactions, skewing the expected partition of elements among its carriers with respect to that expected from thermodynamics.} %\textcolor{black}{The mainline VULCAN runs have been integrated for $10^9$ seconds.}
%As for the starting elemental abundances, we adopted the Solar nebula ones in \cite{asplund21}.
%The CK chemical network only includes H, He, C, N, O, Mg, Si, S and Fe. Consequently, while the CE model can trace also other elements (Al, Na, Ca etc.), we set their abundance to zero to make the results comparable.

Results \textcolor{black}{below} are discussed both in terms of the fraction of element $Y$ carried by each chemical species $X$, $[X]_Y$, and in terms of VMR of the species $X$ in a given atmospheric layer, $[X]=N_X/N_{tot}$. \textcolor{black}{The two atmospheric layers, $10^{-4}$ and 1 bar, have been chosen to discuss our results with respect to those presented in \cite{fegley10} and \cite{visscher10}, respectively. For each element, we plot separately the main and minor carriers to facilitate the reader. The main carriers are those that carry at least 1\% of the tracked element in at least one temperature point, i.e. for which $[X]_{E}\ge10^{-2}$ with $E \in \{\text{Mg},\, \text{Si},\, \text{Fe},\, \text{O}\}$.}
%\textcolor{black}{The two atmospheric layers at which the partitions are shown, 0.1 mbar and 1 bar, represent the extremes of the regions usually probed by spectroscopical surveys \citep[e.g.][]{greene16,tinetti18}.}

\begin{figure*}
\centering
\begin{minipage}[c]{0.80\textwidth}
\centering
%\isPreprints{\centering}{} % Only used for preprints
\includegraphics[width=0.49\textwidth]{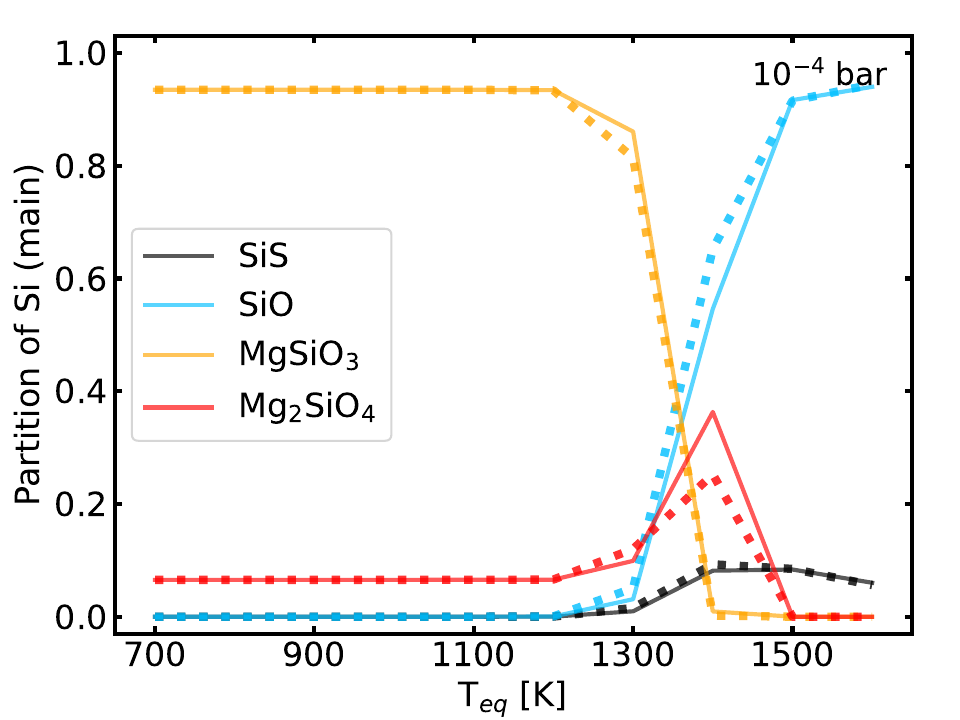}
\includegraphics[width=0.49\textwidth]{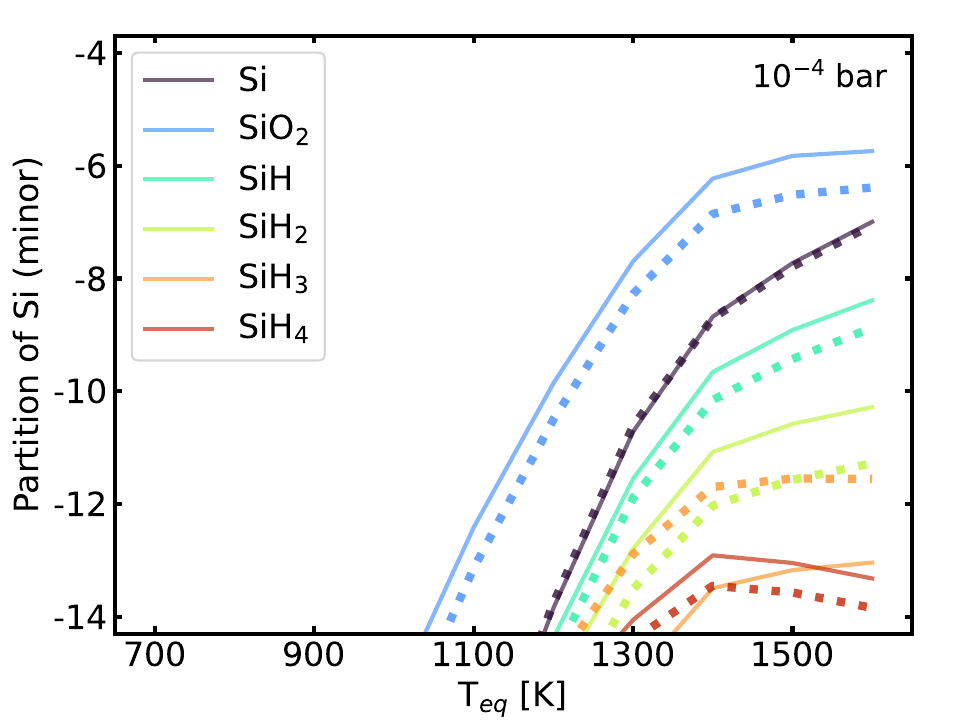}\\
\includegraphics[width=0.49\textwidth]{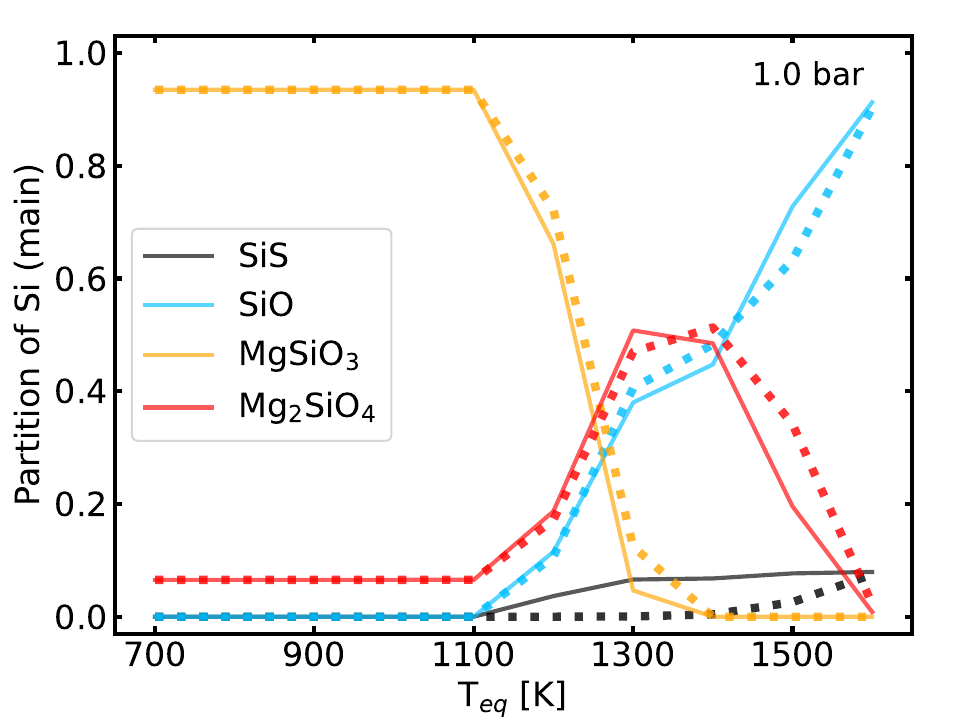}
\includegraphics[width=0.49\textwidth]{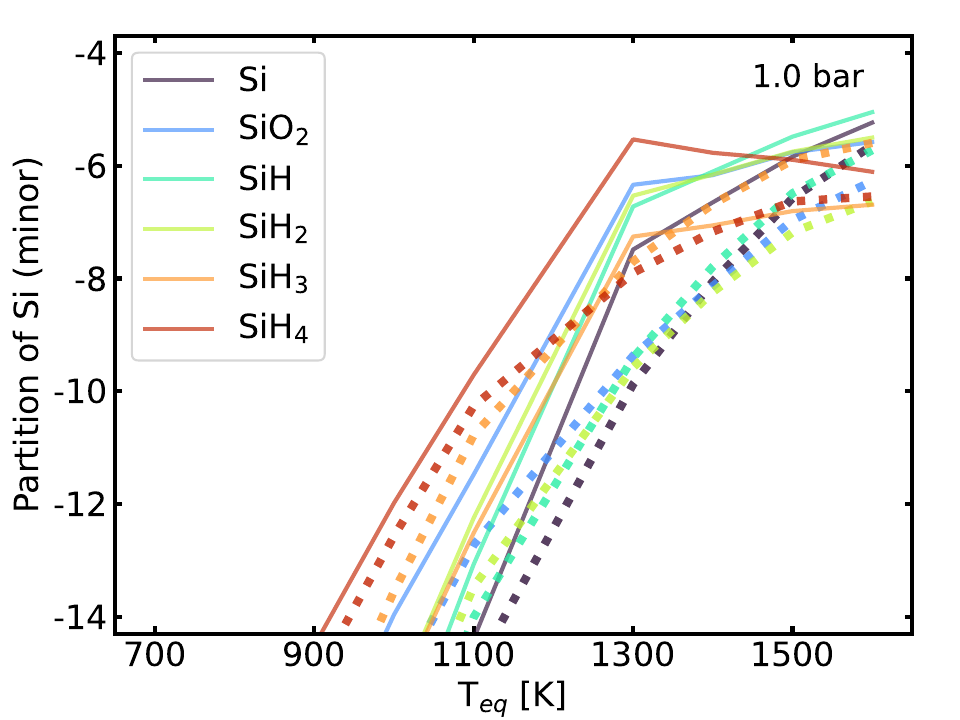}\\
\end{minipage}%
\hfill
\begin{minipage}[c]{0.19\textwidth}
\small
\caption{The partition of Si, $[X]_{Si}$, in its main (left, $[X]_{Si} \ge 0.01$) and minor (right, $[X]_{Si} < 0.01$) carriers at the $10^{-4}$ bar (upper) and 1 bar (lower) levels within the atmosphere, as calculated by the CK model VULCAN (solid) and the CE model FastChem3 (dotted) in the [700, 1600] K $T_{eq}$ interval. Notice that the y-axes of panels on the left are linear, while on the right are logarithmic. \textcolor{black}{Only MgSiO$_3$ and Mg$_2$SiO$_4$ are condensed species, while all the other are in the gas phase.}\label{fig:Si_partition}}
\end{minipage}
\end{figure*}

\subsection{Si partition}
\label{ss:Si_partition}

Three regimes, occurring at different $T_{eq}$, can be identified in the behavior of the main Si-carriers (SiS, SiO, MgSiO$_3$ and Mg$_2$SiO$_4$, left panels of Figure \ref{fig:Si_partition}). Those are: (i) the MgSiO$_3$-dominated, (ii) the Mg$_2$SiO$_4$ peak and (iii) the SiO-dominated ones. The same regimes are found both in the CK and in the CE model \textcolor{black}{at both heights considered, confirming that \texttt{SILANE} is indeed capable of reproducing the main features of the Si partition in the analyzed temperature range. The most notable differences in the main Si carriers between CE and CK calculations are the overproduction of SiS in the [1200, 1500] K range and of SiO at $T_{eq} \ge 1500$ K, at the 1-bar level in the atmosphere. As far as the minor Si carriers is concerned, the CK model tends to overproduce all the tracked species when $T_{eq}$ is between 1200 and 1300 K in the lower atmospheric layer (bottom right panel). This corresponds to the $T_{eq}$ value at which MgSiO$_3$ disappears, and it is likely related to the release of a larger quantity of Si in the CK code with respect to the CE calculations due to the simplified condensation treatment adopted. At the $10^{-4}$ bar layer, where major carriers in the CK model are in better agreement to the CE one, the differences in the minor carriers are also smaller.}

\subsection{Mg and Fe partitions}
\label{ss:MgFe_partition}

\begin{figure*}
\centering
%\isPreprints{\centering}{} % Only used for preprints
\begin{minipage}[c]{0.80\textwidth}
\centering
\includegraphics[width=0.49\textwidth]{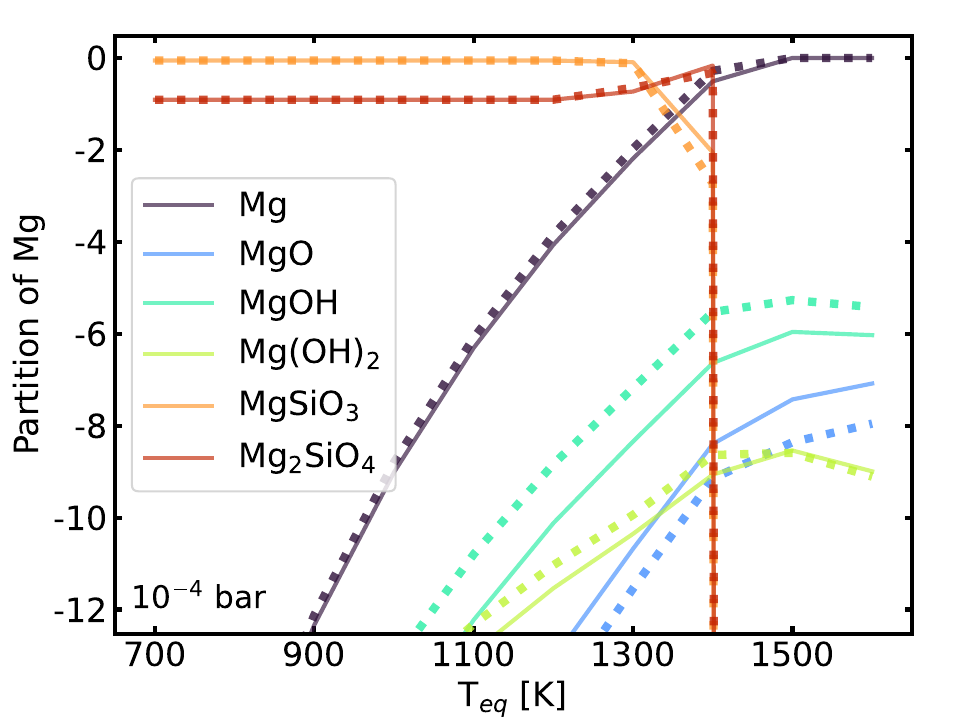}
\includegraphics[width=0.49\textwidth]{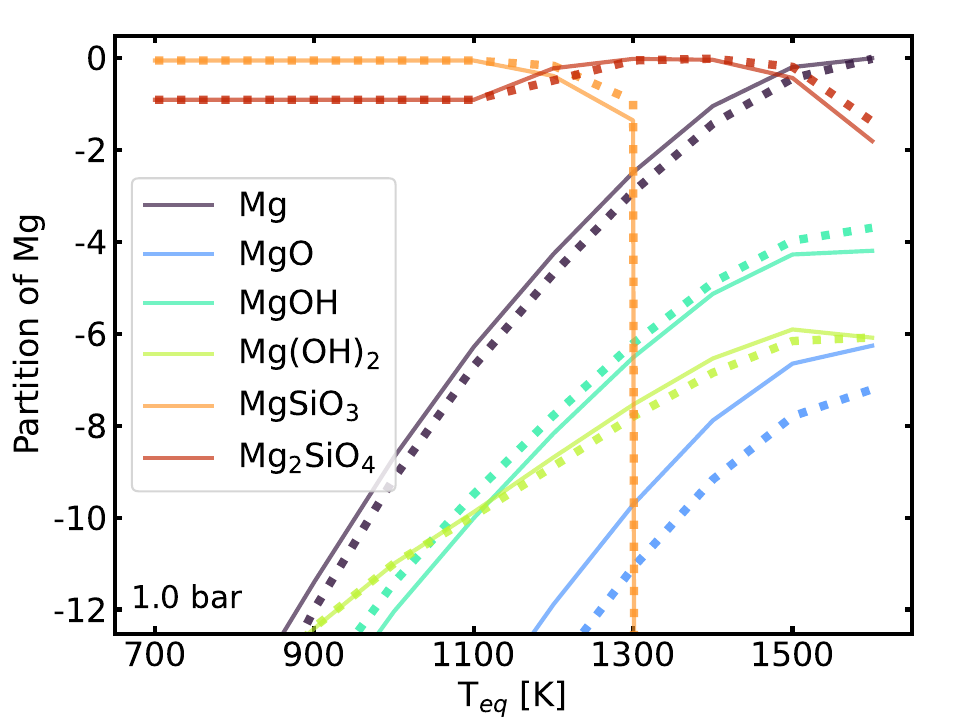}\\
\includegraphics[width=0.49\textwidth]{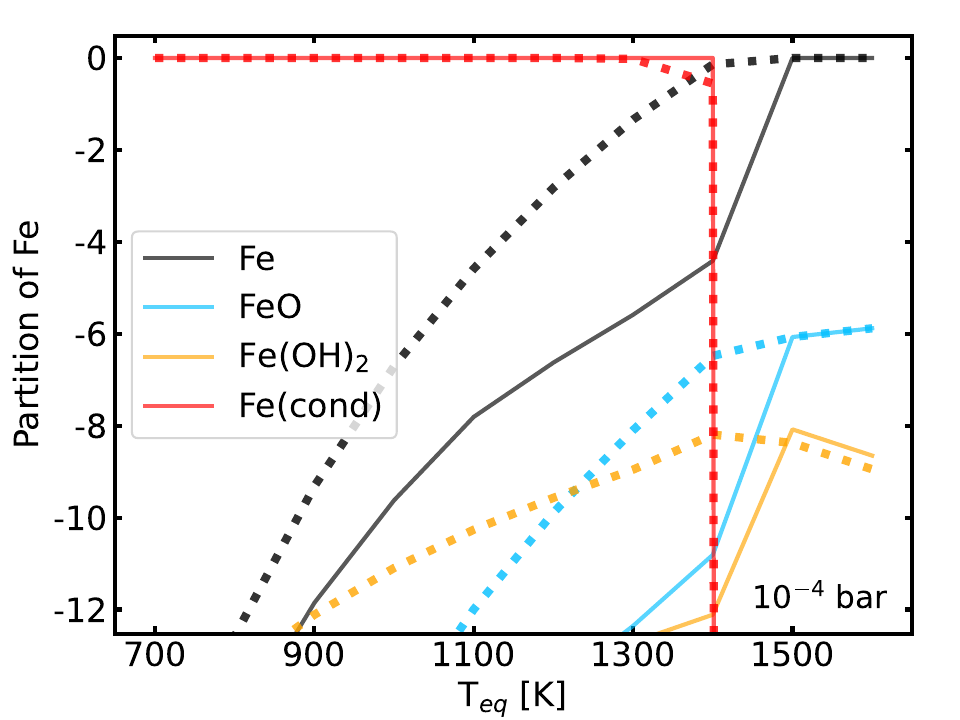}
\includegraphics[width=0.49\textwidth]{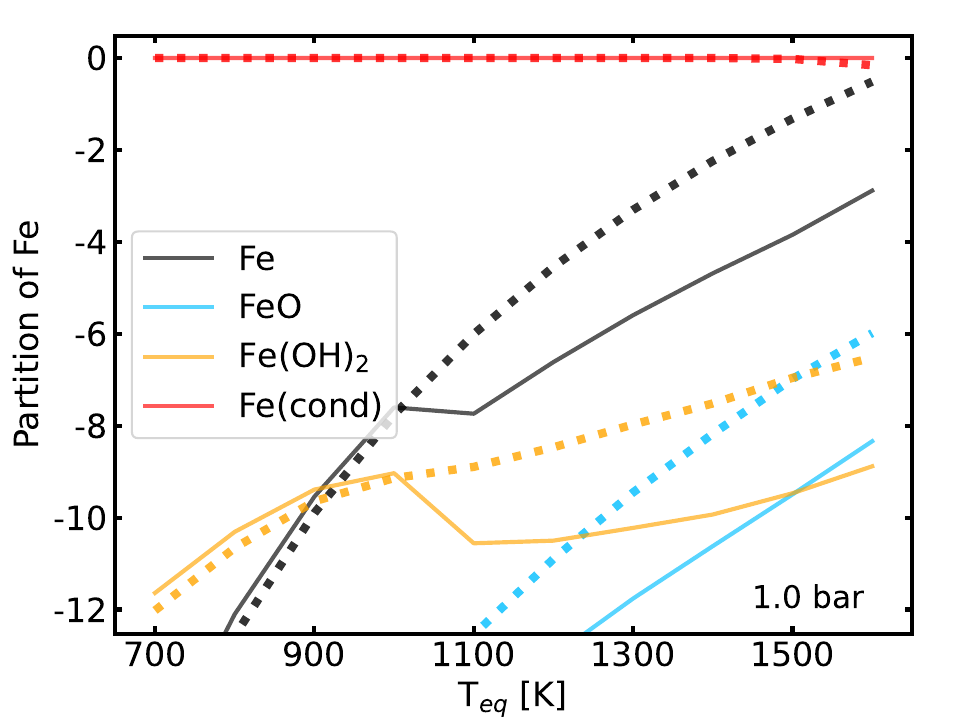}\\
\end{minipage}%
\hfill
\begin{minipage}[c]{0.19\textwidth}
\small
\caption{The partition of Mg ($[X]_{Mg}$, upper) and Fe ($[X]_{Fe}$, lower) in their carriers at the $10^{-4}$ bar (left) and 1 bar (right) levels within the atmosphere, as calculated by the CK model VULCAN (solid) and the CE model FastChem3 (dotted) in the [700, 1600] K $T_{eq}$ interval. \textcolor{black}{Only the species marked as Fe(cond), MgSiO$_3$ and Mg$_2$SiO$_4$ are condensed, while all the others are in the gas phase.} \label{fig:MgFe_partition}}
\end{minipage}
\end{figure*}

The upper panels of Figure \ref{fig:MgFe_partition} shows the partition of Mg among its carriers, again at the $10^{-4}$ and 1 bar levels. The curves of the three main ones, MgSiO$_3$, Mg$_2$SiO$_4$ and Mg, mirror those in Figure \ref{fig:Si_partition}, since the species involved are the same and the starting abundances of Mg and Si are similar. The partition of Mg is thus dominated by MgSiO$_3$, Mg$_2$SiO$_4$ and elemental Mg in the lower, middle and upper parts, respectively, of the explored temperature interval at both atmospheric depths. \textcolor{black}{As seen for Si, both the nature of the two regimes and the transition temperature are the same in the CK and CE model. The correspondence in the partition of Mg, even in minor Mg carriers, is better than in the case of Si, and most differences are $\le$1 order of magnitude.}
%As seen for Si, the transitions between different regimes are shifted to lower temperatures by 100-200 K with respect to CE and the endmember $[X]_{Mg}$ are equivalent in the CK and the CE cases within 4\%. The presence of the virtual gas phases of MgSiO$_3$ and Mg$_2$SiO$_4$ in the CK model is shown explicitly here, since in place of the $[X]_{Mg}$ cliff of the CE model (dashed line), there is a gentler decrease of their partition down to around $10^{-6}$ of the total count of Mg atoms. The behavior of the minor Mg-carriers MgO, MgOH and Mg(OH)$_2$ is also reproduced, however their $[X]_{Mg}$ is up to $\sim$3 orders of magnitude higher at intermediate temperatures, and $\sim$1 order of magnitude lower at 1600 K at the 1 bar level (right panel of Figure \ref{fig:MgFe_partition}). Despite these differences, their atmospheric mixing ratios remain below $10^{-8}$. As for the Si partition, the overall partition of Mg is not impacted by the presence of the virtual gaseous enstatite and forsterite.

The lower panels of Figure \ref{fig:MgFe_partition} shows the partition of Fe among its CK and CE largest-fraction carriers for the same two atmospheric layers as previous Figures, i.e. $10^{-4}$ and 1 bar. Differently from rock-forming silicates, the iron partition is dominated by elemental Fe. Pure iron is mostly locked in solid or molten droplets (depending on the temperature) aside for the high-end tail of the investigated $T_{eq}$ interval, where it is in gaseous phase. The transition between the two regimes is sharp as it is driven by condensation and evaporation (left panel of Figure \ref{fig:MgFe_partition}). The disappearance of the condensed iron phase (red curve) happens at an higher temperature in the CK model. This is related to a small difference in the condensation curve prescription that we used (see Eq.~\ref{eq:Fe_condensation}) with respect to the one adopted by FastChem3. The vertical profile (not shown here) shows that the condensation curve crosses the pressure-temperature profile at a slightly higher altitude, thus allowing for the presence of solid-state Fe at the $10^{-4}$ bar level. As for the abundances of the other Fe-bearing species, \textcolor{black}{they are generally lower than in the CE calculations, again due to a slightly lower estimate of the Fe vapor pressure in the CK model, that suppresses the availability of Fe atoms in the gas phase.}
%As for the abundances of the other Fe-bearing species, they are generally higher in the CK calculations than in the CE one, up to $\sim$ 4 orders of magnitude for Fe(OH)$_2$. This is likely due to the fact that our chemical network does not trace FeS, since we were not able to find any published reaction rate for the formation of this molecule in the gas phase. In the CE model, at the $10^{-4}$ bar level, FeS is found either as a condensate at $T_{eq} = 700$ K, where it locks up 26\% of the available iron, or as a gas at $T_{eq} \ge 1500$ K, where it stores around $10^{-4}$ of all the iron atoms ([FeS]$\sim3\times 10^{-9}$). However, all these differences involve species that sequester very low amounts of iron (apart for the solid FeS at 700 K), thus producing essentially negligible differences between the CK- and CE-derived atmospheric chemical makeups in the investigated temperature range.

\begin{figure*}
\centering
\begin{minipage}[c]{0.80\textwidth}
\centering
%\isPreprints{\centering}{} % Only used for preprints
\includegraphics[width=0.49\textwidth]{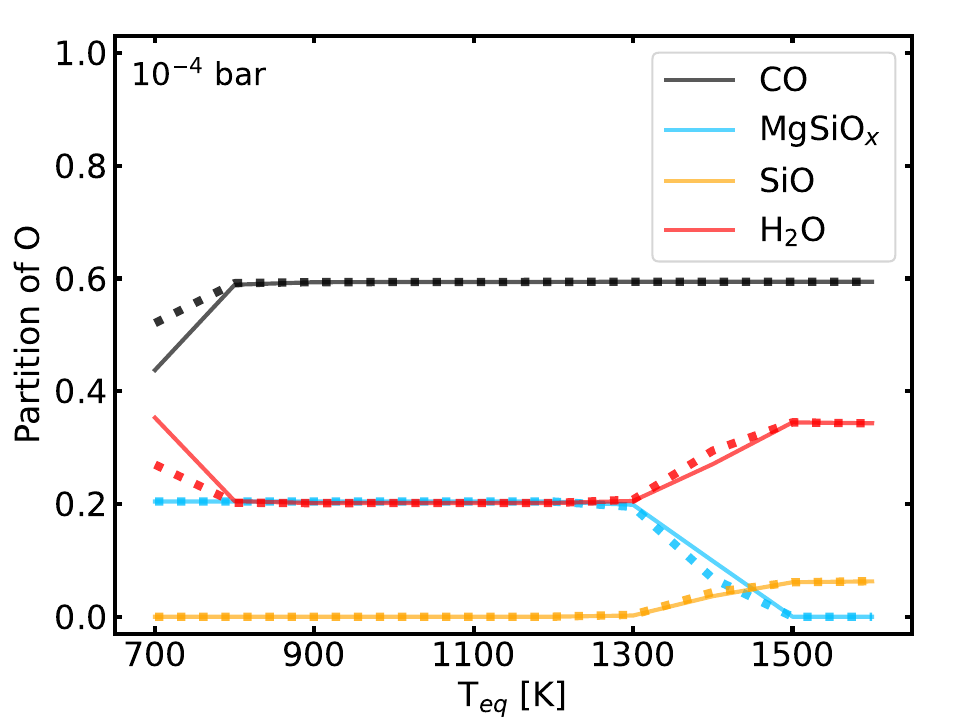}
\includegraphics[width=0.49\textwidth]{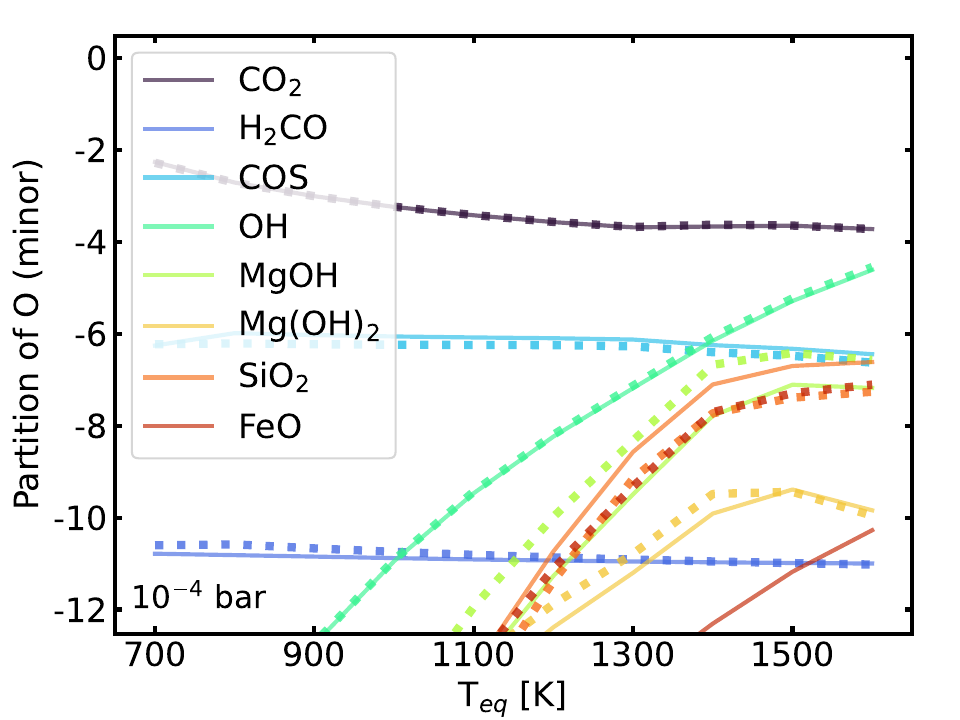}\\
\includegraphics[width=0.49\textwidth]{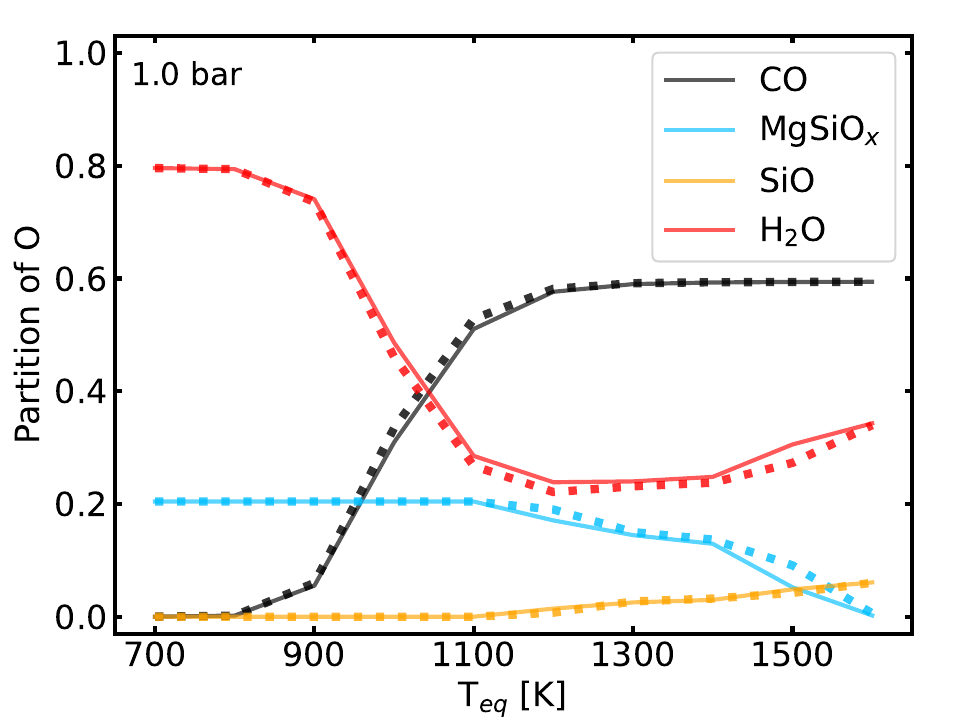}
\includegraphics[width=0.49\textwidth]{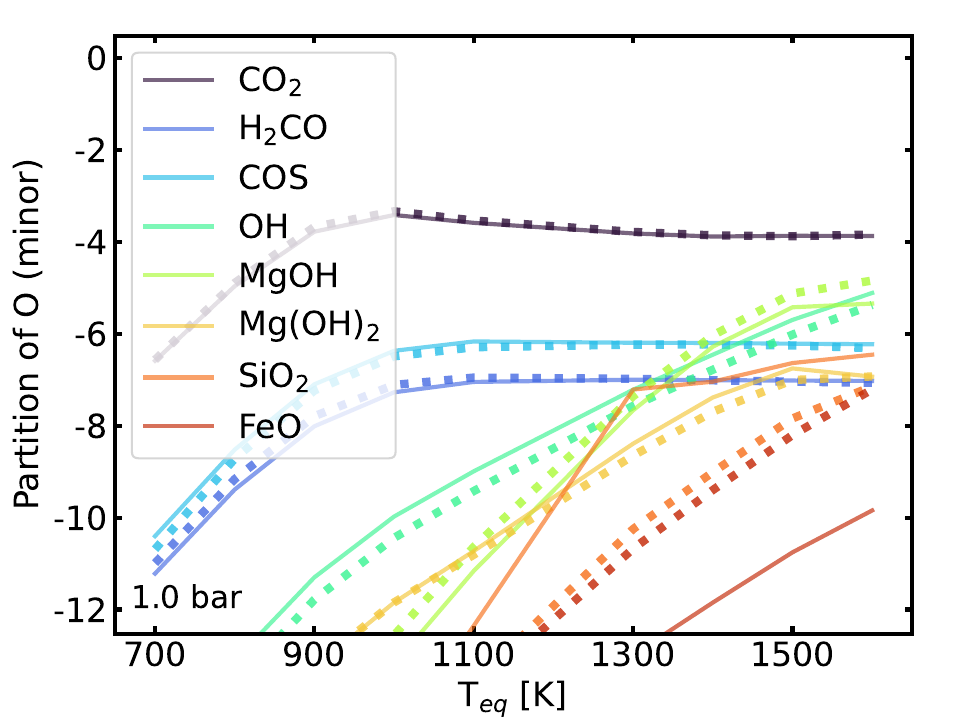}\\
\end{minipage}%
\hfill
\begin{minipage}[c]{0.19\textwidth}
\small
\caption{The partition of O, $[X]_O$, in its main (left) and minor (right) carriers at the $10^{-4}$ bar (upper) and 1 bar (lower) levels within the atmosphere, as calculated by the CK model VULCAN (solid) and the CE model FastChem3 (dotted) in the [700, 1600] K $T_{eq}$ interval. Notice that the y-axes of panels on the left are linear, while on the right are logarithmic. \textcolor{black}{Only MgSiOx, which represents the sum of the oxygen fractions locked into enstatite and forsterite, is not in the gas phase.}\label{fig:O_partition}}
\end{minipage}
\end{figure*}

\subsection{O partition}
\label{ss:O_partition}

Finally, Figure \ref{fig:O_partition} shows the partition of oxygen among major (left) and minor (right) species at the $10^{-4}$ (upper) and $1.0$ (lower) bar levels. While oxygen is not a refractory, its partition is worth studying since some of its carriers (e.g. H$_2$O and CO) are \textcolor{black}{usually abundant enough to be observed} in exoplanetary atmospheres, and their VMR form the basis to calculate the elemental abundances of O and C. \textcolor{black}{In turn, these abundances can be used to reconstruct the formation and evolution history of the planet \citep[][]{oberg11,turrini21,pacetti22}}. \textcolor{black}{Determining the amount of oxygen not available to form H$_2$O and CO is particularly interesting, since this unseen component can skew the C/O ratio.} In the plots, the O shares of MgSiO$_3$ and Mg$_2$SiO$_4$ are summed together and marked as MgSiOx (light blue line). As expected, the main O carriers are H$_2$O and CO, which are the most abundant at lower and higher $T_{eq}$, respectively. The exact $T_{eq}$ where CO overtakes H$_2$O depends on the position in the atmosphere, but both the CK and the CE models agree on the same value. Three other important O-bearing species are MgSiO$_3$ and Mg$_2$SiO$_4$ at the low-end tail and SiO at high-end one of the $T_{eq}$ interval. Specifically, in our Solar composition atmosphere, MgSiOx captures around 20\% of all the O atoms at 700 K, while SiO carries up to 6\% of oxygen at 1600 K. The partition curves are \textcolor{black}{nearly identical in the two models, as expected from the previous results on the partition of Si and Mg. The same is true also for the minor O carriers, which are all in good agreement apart for SiO$_2$ and FeO, both of which can differ up to three orders of magnitude with respect to their fractional abundance evaluated at CE.} However, \textcolor{black}{it is worth noticing that their} impact on the general partition of O is small, since it involves molecules whose $[X]_{O}$ is $\sim$$10^{-6}$, \textcolor{black}{which translates to a VMR of $\sim$10$^{-10}$ when accounting for the oxygen abundance.}
%thus around $\sim0.1$ \textcolor{black}{ppbv} at maximum in absolute mixing ratios.
%The partition curves are similar between CE and CK, yet transitions between different regimes occur about 200 K earlier in CK, which is the same behavior discussed in the partition of Si (Sect.~\ref{ss:Si_partition}) and Mg (Sect.~\ref{ss:MgFe_partition}).
%In the CE model, there is a distinct minimum in the H$_2$O abundance at around 1300 K, where CO is already the dominant carrier but silicates are still present. This minimum disappears in the CK model, and excess oxygen freed by MgSiOx dissociation ends up bonded in H$_2$O. The trends of the minor O carriers (Figure \ref{fig:O_partition}, right panels) are similar in the CE and CK models, especially for the relatively abundant ones (CO$_2$, OH and COS), while they diverge at intermediate temperatures for the others. However, the impact on the general partition of O is small, since it involves molecules whose $[X]_{O}$ is $\sim 10^{-5}$, thus around few pptv at maximum in absolute mixing ratios.

%%%%%%%%%%%%%%%%%%%%%%%%%%%%%%%%%%%%%%%%%%%%%%%%%%%%%%%%%%%%%%%%%%
%%%%%%%%%%%%%%%%%%%%%%%%%%%%%%%%%%%%%%%%%%%%%%%%%%%%%%%%%%%%%%%%%%

\section{Application cases}
\label{sec:results}

\textcolor{black}{In this Section we apply \texttt{SILANE} to two cases: the partition of Si, Mg and oxygen in a vertically mixed Solar-like atmosphere as a function of the planetary equilibrium temperature and the determination of the vertical mixing strength in the case of the $\sim$15 Myr-old Jupiter YSES-1c. These calculations include molecular and eddy diffusion, condensation and sedimentation following the equations shown in Sect.~\ref{sec:model}. We assume a constant value for the eddy diffusion coefficient over the entire atmosphere. As in Section \ref{sec:comparison}, condensed particle diameters are set to 1 $\mu$m, a zero-flux boundary closure is adopted and photochemistry is not taken into account.}

\subsection{A hot Jupiter-like planet}
\label{ss:hotjup}

\begin{figure*}
\centering
\begin{minipage}[c]{0.80\textwidth}
\centering
%\isPreprints{\centering}{} % Only used for preprints
\includegraphics[width=0.42\textwidth]{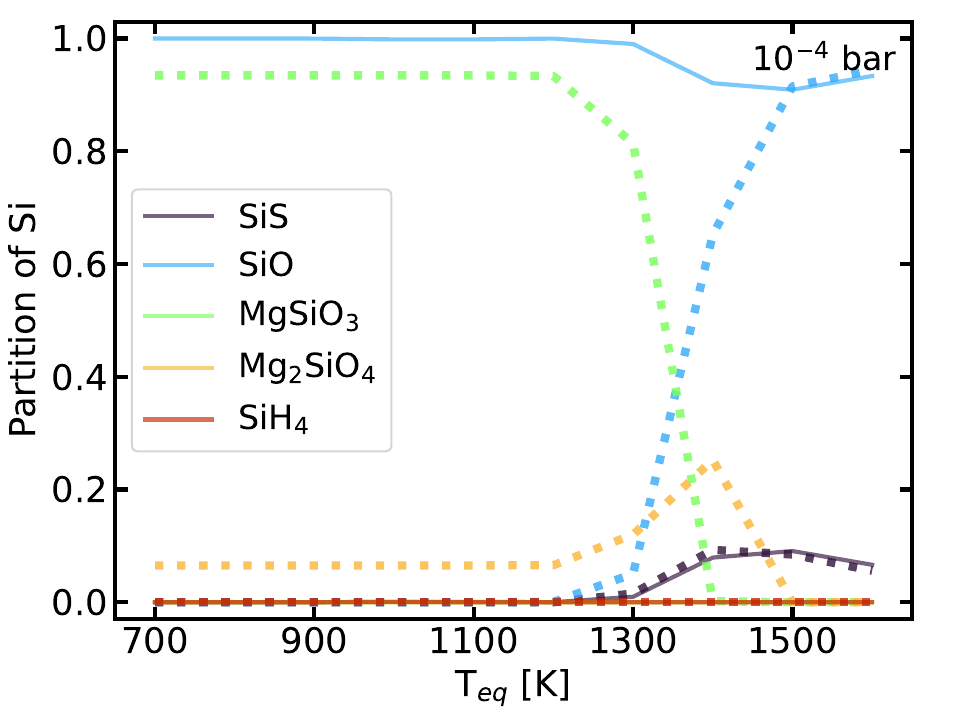}
\includegraphics[width=0.42\textwidth]{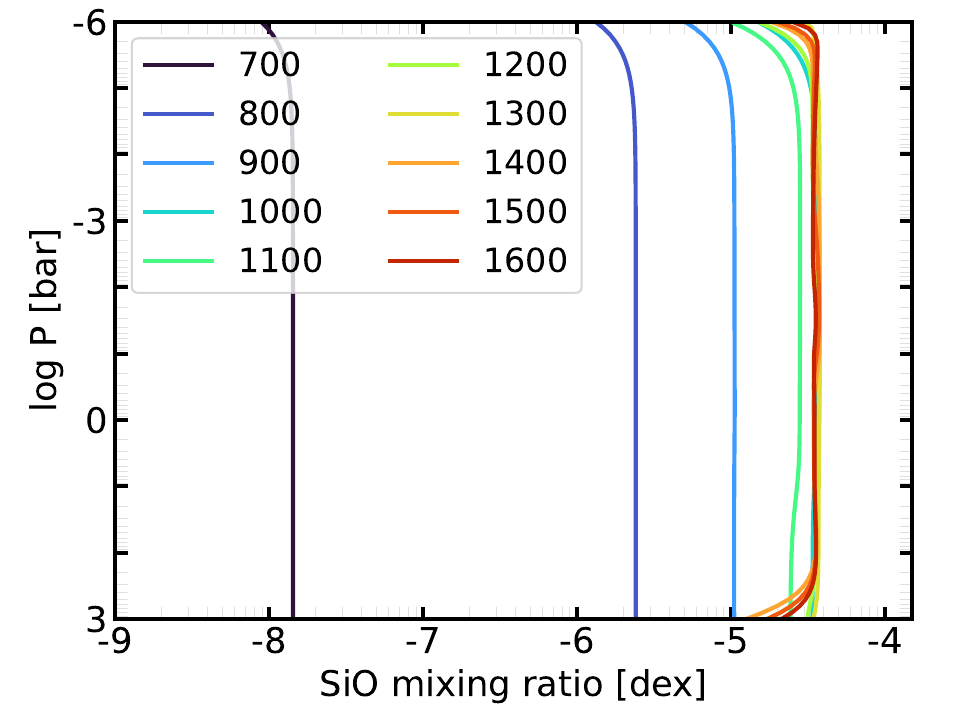} \\
\includegraphics[width=0.42\textwidth]{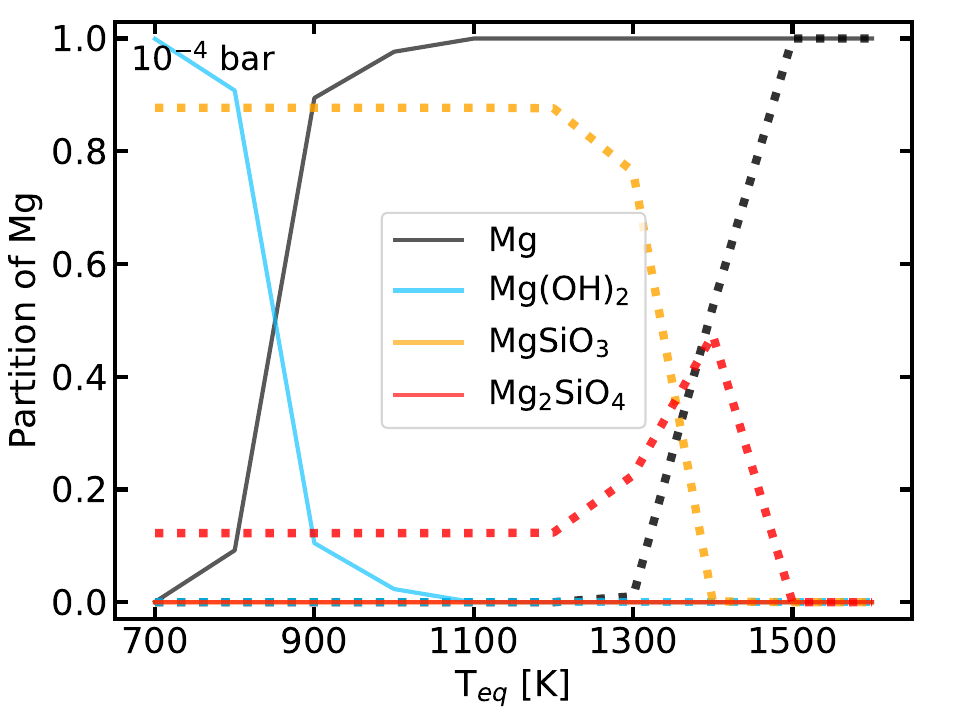}
\includegraphics[width=0.42\textwidth]{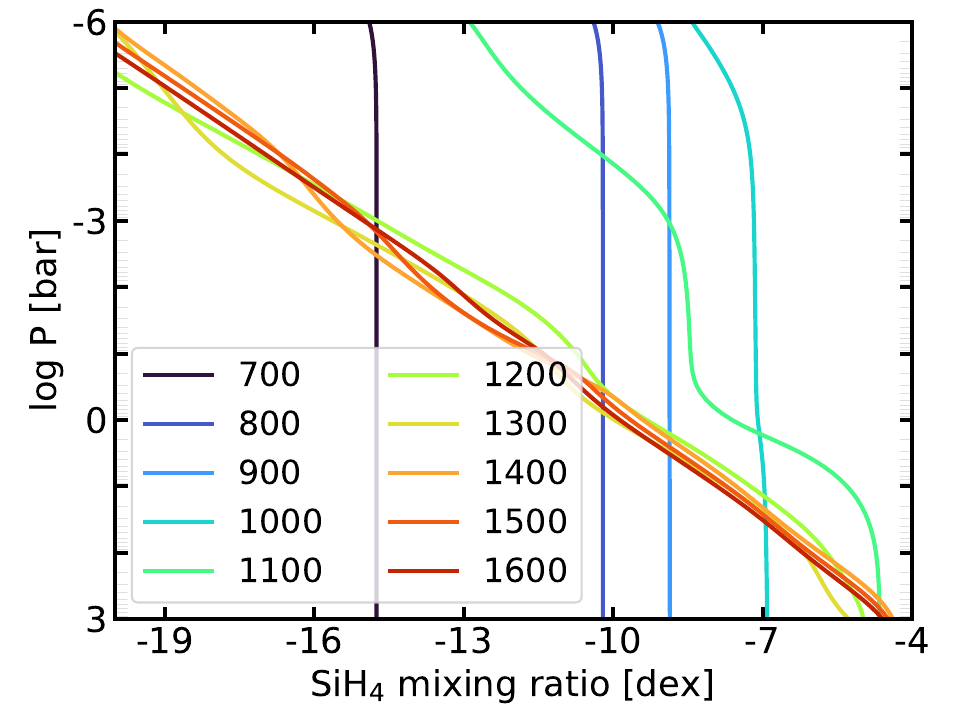} \\
\includegraphics[width=0.42\textwidth]{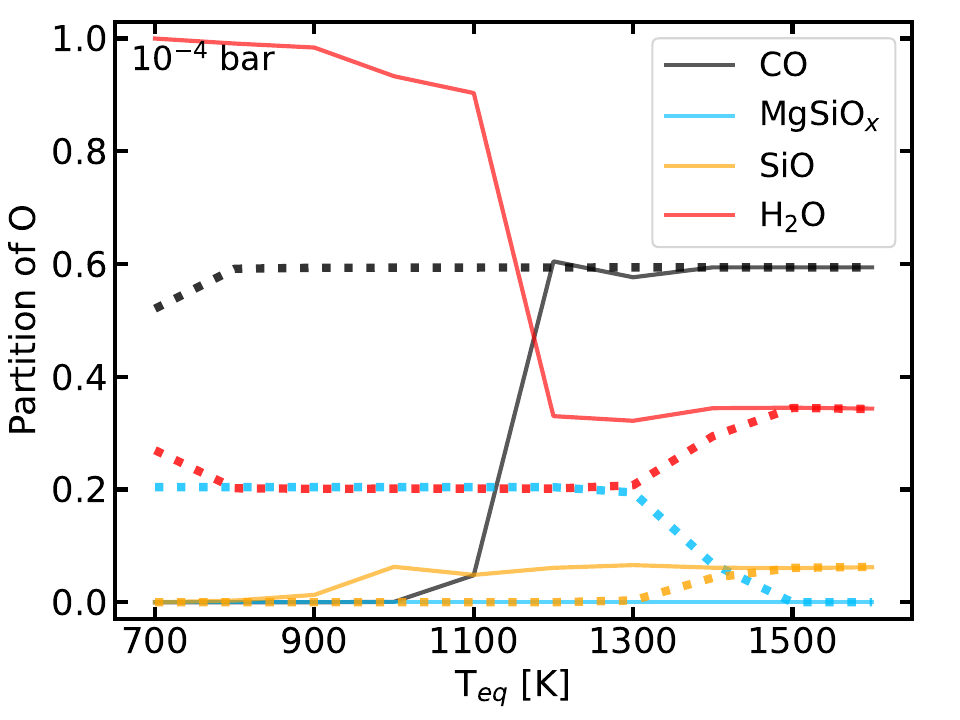}
\includegraphics[width=0.42\textwidth]{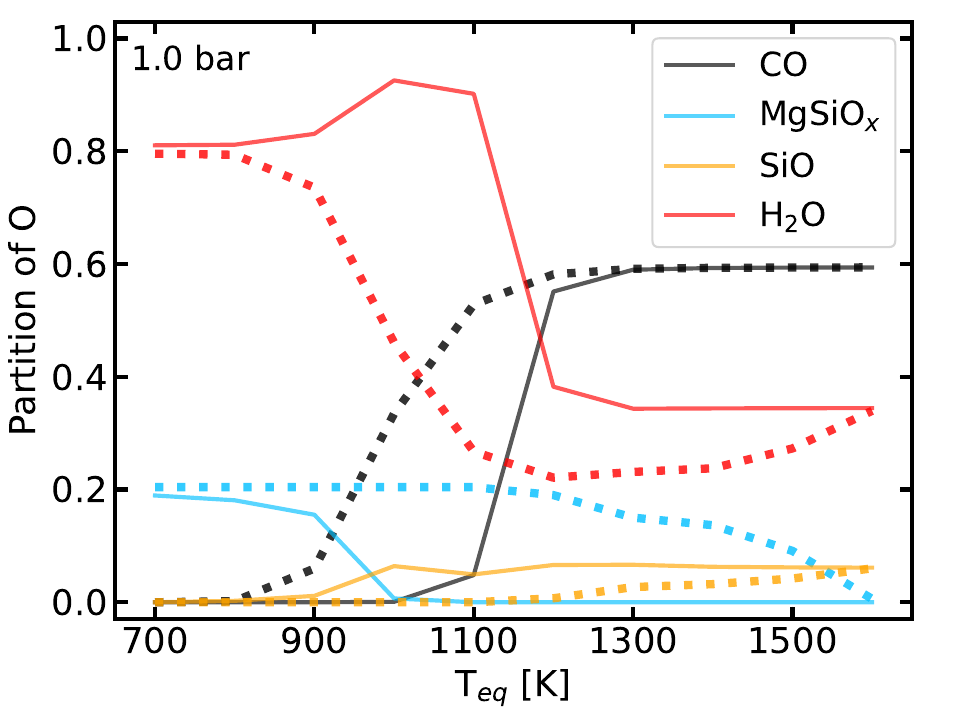}\\
\end{minipage}%
\hfill
\begin{minipage}[c]{0.19\textwidth}
\small
\caption{\textcolor{black}{Chemical models of a Jupiter-like extrasolar planet with a Solar-composition, vertically mixed atmosphere. Left column: elemental partitions of Si, Mg and O at the $10^{-4}$ bar level. Right column: vertical VMR profiles for SiO and SiH$_4$ at different $T_{eq}$, plus the O partition at the 1 bar level.} Solid lines are the CK results. Dotted lines are the CE results. \label{fig:Atmo_mixing}}
\end{minipage}
\end{figure*}
%Upper left: partition of Si, $[X]_{Si}$. Upper right: the vertical distribution of SiH$_4$ at different $T_{eq}$. Lower left: partition of Mg, $[X]_{Mg}$. Lower right: partition of O , $[X]_{O}$.

\textcolor{black}{For this analysis, we set the $K_{zz}=10^{8}$ cm$^2$s$^{-1}$ and determine the O and Si partition at two levels within the atmosphere: 1 and $10^{-4}$ bar.}
%After having characterized our new CK silicate network in equilibrium conditions, we turn on again the diffusion term in Eq.~\ref{eq:vulcan} to study the impact of vertical dynamics on the abundances of Mg, Si, Fe and O carriers in a Solar-composition exoplanetary atmosphere. Specifically, our calculations include molecular and eddy diffusion, condensation and sedimentation following the equations shown in Sect.~\ref{sec:model}. The eddy diffusion coefficient has a single value across the entire atmosphere, set to $10^{8}$ cm$^2$s$^{-1}$. Condensed particle diameters are set to 1 $\mu$m and the boundaries of the atmospheric column are closed, i.e. there is no exchange of mass.
As before, we model our planet after Jupiter and consider $T_{eq}$ values in the [700, 1600] K range. VULCAN is left running until convergence is achieved or at least $10^7$ seconds have lapsed since the start of the simulation. A more thorough investigation, involving a wider set of metallicities and eddy diffusion values, is performed in a companion paper \citep[][]{simonetti26}.

The results for the $10^{-4}$ bar level are shown in Figure \ref{fig:Atmo_mixing}\textcolor{black}{, left column}. The inclusion of vertical mixing has several different impacts on the partition of the tracked elements. As far as the Si and Mg partition is concerned, we identify two main results. \textcolor{black}{First, there are essentially no MgSiO$_3$ and Mg$_2$SiO$_4$ within this temperature range. This is expected, since our 1 $\mu$m grains would need a pressure $\sim100$ larger to stay afloat. The main Si and Mg carriers are, respectively, SiO and either Mg(OH)$_2$ (for $T_{eq}<900$ K) or Mg. While the partition fraction of SiO never falls below $\sim90$\%, its actual VMR vary substantially with $T_{eq}$, increasing from $10^{-8}$ at 700 K to $3\times10^{-5}$ at 1600 K (Figure \ref{fig:Atmo_mixing}, upper right panel, light blue curve). Similarly, the relative importance of Mg(OH)$_2$ does not translates directly into an high VMR of this species, since it increases from $10^{-8}$ at 700 K to $2\times10^{-6}$ at 800 K, and then decreases again for higher temperatures (not shown). The low VMR of both SiO and Mg(OH)$_2$ at the lower end of the $T_{eq}$ interval is a product of the vertical sequestration of Mg and Si that removes these species from the $10^{-4}$ bar layer into a silicate cloud deck deeper within the atmosphere. The presence of this layer is underlined by the partition of O at the 1 bar level (Figure \ref{fig:Atmo_mixing}, lower right panel), for $T_{eq} < 1000$ K.} Second, a small fraction of Si freed from silicates dissolution forms SiH$_4$ at a substantially larger rate than in the equilibrium case in the [900, 1100] K equilibrium temperature range, with peak VMR values in the upper atmosphere equal to $10^{-7}$ in absence of photodissociation (upper right panel). This might offer the opportunity to directly probe the Si inventory of the atmosphere if successfully retrieved from observations.

The oxygen partition (Figure \ref{fig:Atmo_mixing}, third row) changes substantially with respect to cases without vertical mixing. First of all, the crossover temperature at which CO becomes more abundant than H$_2$O moves \textcolor{black}{up at both atmospheric levels. In particular, at $10^{-4}$ bar, it moves from below} 700 K in the CE case to 1200 K in the CK case \textcolor{black}{In general,} water blocks a larger share of O across the entire $T_{eq}$ range, \textcolor{black}{while} the share of O blocked in refractories starts decreasing at lower $T_{eq}$ and becomes negligible at 1000 K in the CK model, while it remains stable at 20.6\% up to 1300 K in the CE model. In the CK case, SiO starts accumulating just above 900 K and collects up to 6.1\% of the available oxygen in the 1000-1600 K interval. On the other hand, no SiO appears in CE calculations until $T_{eq} > 1300$ K. \textcolor{black}{This substantially changes how the molecular VMR are translated into elemental abundances, since the fraction of oxygen generally considered locked up in refractories is much smaller in a mixed atmosphere than in an equilibrium one in the [900, 1300] K interval.}
%By comparing Figure \ref{fig:Atmo_mixing} with Figure \ref{fig:O_partition}, it is immediately clear that this is an effect of the mixing, rather than of the chosen reaction rates. This means that, for different $K_{zz}$ values, we expect both the H$_2$O/CO crossover temperature and the dependence on temperature of the fraction of oxygen sequestered by refractories to change.

%Finally, we show the vertical mixing ratio profiles of SiH$_4$, at different $T_{eq}$. In the CE model, SiH$_4$ is a minor Si carrier with very low mixing ratios in the observable atmosphere. Part-per-million concentrations are reached near the very bottom of our simulated column (1000 bar) and only for the highest $T_{eq}$ value tested (1600 K). The non-mixed CK model produces higher amounts of SiH$_4$ and, differently from the CE model, the highest concentrations are found for $T_{eq} = 900$ K. However, [SiH$_4$] is still below $10^{-6}$ when $P < 10$ bar even in the most favorable case. On the other hand, in atmospheres with mixing (Figure \ref{fig:Atmo_mixing}, lower right panel), SiH$_4$ has a much higher concentration up to the top of the atmosphere. In particular, the cases with $T_{eq}$ equal to 800, 900 and 1000 K all sport [SiH$_4$] values in the 1-3$\times10^{-5}$ with little vertical variations, which may imply the observability of this species even in cloudy atmospheres.

%%%%%%%%%%%%%%%%%%%%%%%%%%%%%%%%%%%%%%%%%%%%%%%%%%%%%%%%%%%%%%%%%%
%%%%%%%%%%%%%%%%%%%%%%%%%%%%%%%%%%%%%%%%%%%%%%%%%%%%%%%%%%%%%%%%%%

\subsection{YSES-1 c}
\label{ss:yses1c}

\begin{table}
\centering
\caption{Planetary and atmospheric parameters of YSES-1c.}
\begin{tabular}{ll}
\hline
Parameter & Value \\
\hline
Gravity & 25.7 m s$^{-2}$ \\
Z/Z$_{\odot}^1$ & 3.3 \\
$\text{[C/H]}^2$ & 0.66 \\
$\text{[O/H]}^2$ & 0.50 \\
$\text{[S/H]}^2$ & 0.81 \\
\textcolor{black}{$r_p$} & \textcolor{black}{0.1 $\mu$m} \\
\hline
k$_{zz}$$^3$ & 10$^4$, 10$^5$, 10$^6$, 10$^7$, 10$^8$, 10$^{9}$, \textcolor{black}{10$^{10}$} cm$^2$ s$^{-1}$ \\
\hline
\multicolumn{2}{l}{\small $^1$Used to scale the abundances of all elements not explicitly}\\
\multicolumn{2}{l}{\small retrieved. $^2$[X/H]=log$_{10}$[([X][H]$_\odot$)/([X]$_\odot$[H])]. $^3$ This}\\
\multicolumn{2}{l}{\small work.}
\end{tabular}
\label{tab:yses1c_setting}
\end{table}

The recent direct imaging of the hot Jupiter YSES-1 c using the James Webb Space Telescope showed the presence of spectral features compatible with MgSiO$_3$-Mg$_2$SiO$_4$ clouds in the $10^{-3}$-$10^{-2}$ bar region \textcolor{black}{with a retrieved grain $r_p\le 0.1$ $\mu$m}, an emission temperature of the planet in the 950-1100 K range and a metallicity $\sim3$ times Solar \citep{hoch25}. The instantaneous separation from the parent solar-type star is 320 AU, which means that essentially all the observed thermal flux is due to the internal heat, thus that the planet is young \citep[][]{bohn20}.

\begin{figure*}
\centering
\begin{minipage}[c]{0.75\textwidth}
\centering
%\isPreprints{\centering}{} % Only used for preprints
\includegraphics[width=0.49\textwidth]{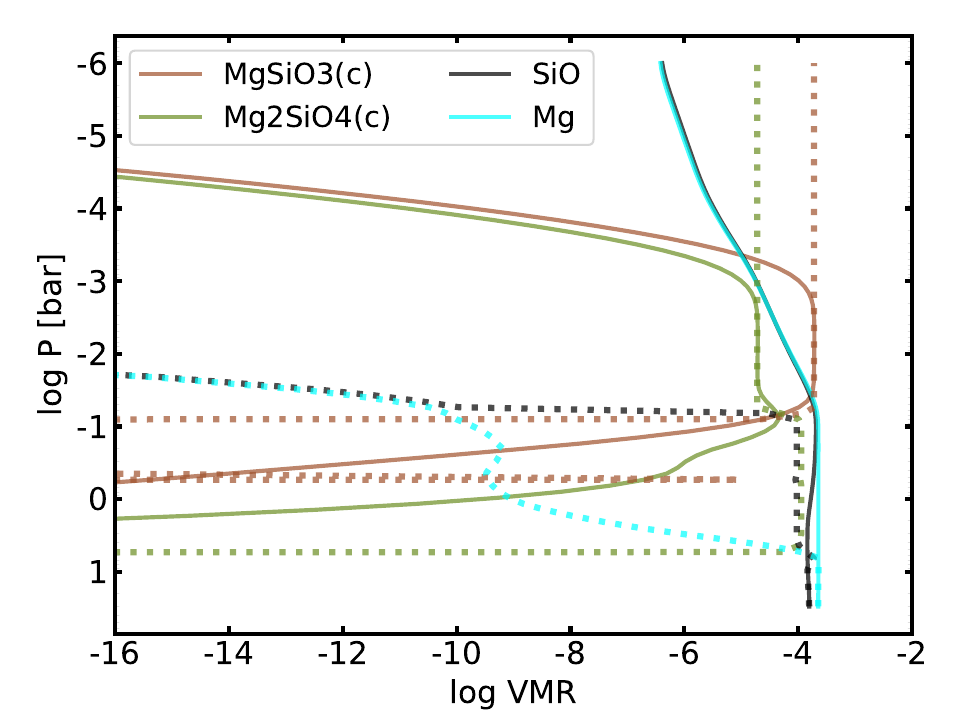}
\includegraphics[width=0.49\textwidth]{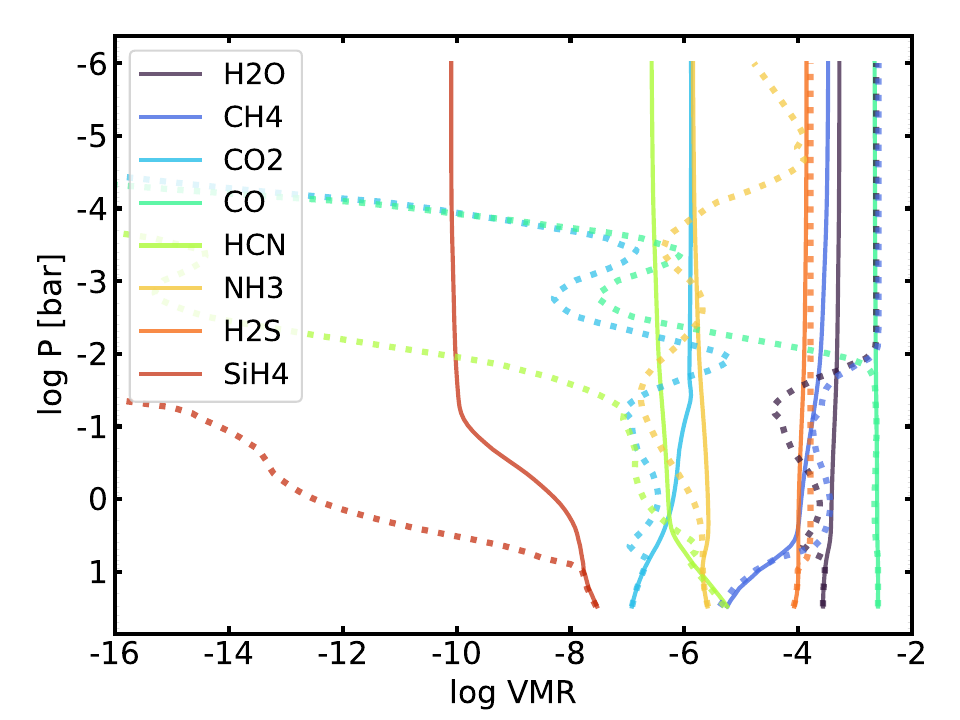} \\
\includegraphics[width=0.49\textwidth]{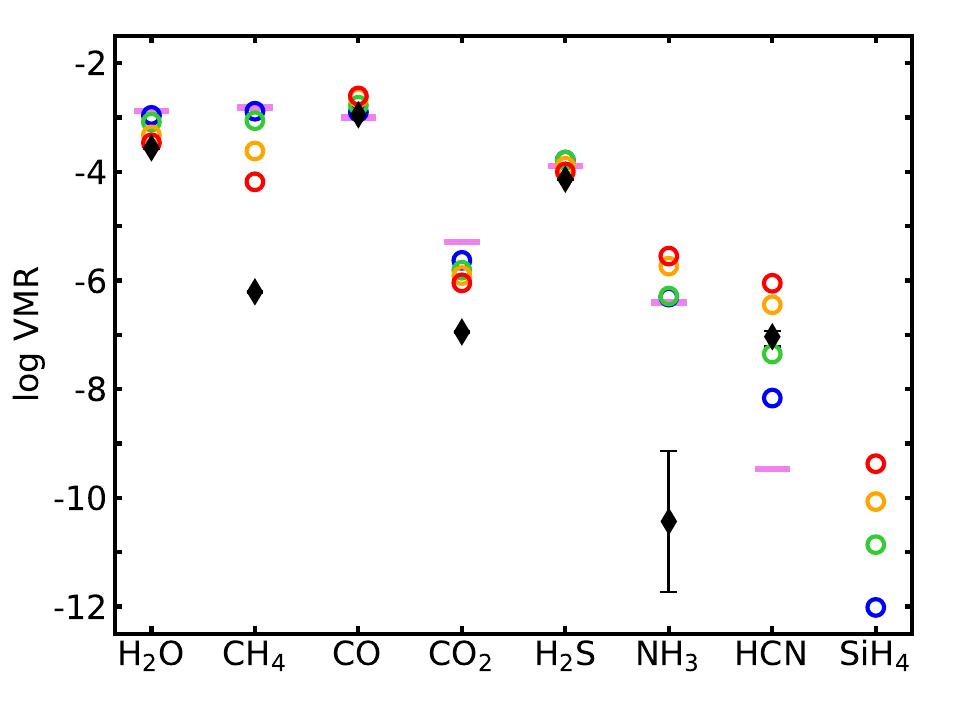}
\includegraphics[width=0.49\textwidth]{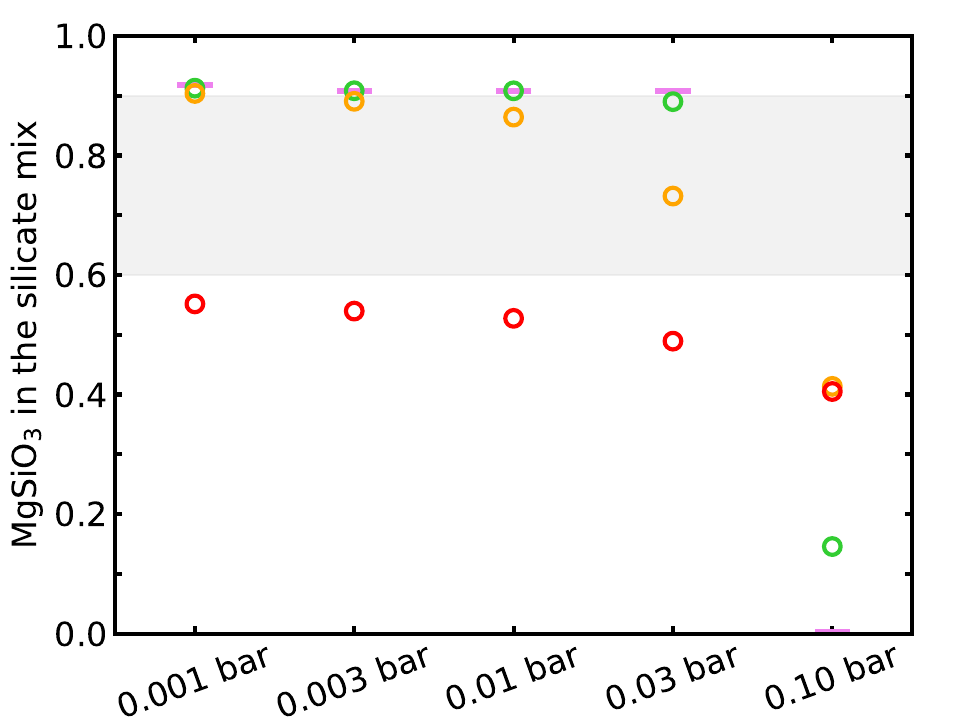}\\
\end{minipage}%
\hfill
\begin{minipage}[c]{0.24\textwidth}
\small
\caption{\textcolor{black}{Chemical makeup of the YSES-1c atmosphere and its dependence on the vertical mixing. Upper left: the vertical VMR profiles for the two silicate condensates and the two main Si and Mg carriers at equilibrium (dotted) and in a mixed atmosphere with $K_{zz}=10^{9}$ cm$^{2}$s$^{-1}$ (solid). Upper right: the vertical VMR profiles for eight volatile chemical species at equilibrium (dotted) and when $K_{zz}=10^{9}$ cm$^{2}$s$^{-1}$ (solid). Lower} left: the integrated VMR of the same eight species in the [0.01, 0.02] bar region. Black diamonds: values retrieved by observations \citep[][]{hoch25}. There is no observational data on the mixing ratio of SiH$_4$. \textcolor{black}{Lower} right: the fraction of MgSiO$_3$ in the \textcolor{black}{condensed} silicates mix at different atmospheric levels. Gray shaded area: retrieved interval. In the two bottom panels, violet horizontal dashes are the equilibrium VMR and empty circles are the disequilibrium VMR for \textcolor{black}{$10^7$ (blue, left panel only), $10^8$ (green), $10^9$ (orange) and $10^{10}$ (red) cm$^2$s$^{-1}$}.\label{fig:yses1c}}
\end{minipage}
\end{figure*}

In order to study this planet, we adopted the retrieved atmospheric, \textcolor{black}{compositional} and planetary parameters retrieved by \cite{hoch25}. We report them in Table \ref{tab:yses1c_setting}. As for the pressure-temperature profile, we adopt the one retrieved by petitRADTRANS\footnote{\cite{molliere19}} and shown as a blue curve in Figure 4 of \cite{hoch25}, \textcolor{black}{clipped at 30 bar, where the predicted temperature is equal to 2830 K. This is sufficient to guarantee both that the bottom of the atmosphere is at chemical equilibrium and that rock-forming silicates evaporate within the column (see Fig.~\ref{fig:yses1c}, upper left panel).} We tested \textcolor{black}{7} different values for the vertical mixing coefficient ($10^4$, $10^5$, $10^6$, $10^7$, $10^8$, $10^9$, \textcolor{black}{and $10^{10}$} cm$^{2}$s$^{-1}$). \textcolor{black}{Given the large instantaneous distance between planet and its parent star at the time of observation, we can safely neglect photochemistry.}

\textcolor{black}{The impact of the vertical mixing on the vertical distributions of the seven volatile species whose abundances have been retrieved (H$_2$O, CO, CO$_2$, CH$_4$, H$_2$S, NH$_3$ and HCN), plus SiH$_4$, is shown in the upper right panel of Figure \ref{fig:yses1c}. The bottom left panel shows instead the integrated VMR in the [0.01, 0.02] bar region in chemical equilibrium (horizontal dashes) and disequilibrium (circles).} The atmospheric region has been selected because most of the planetary emission comes from this layer. \textcolor{black}{As such, the mixing ratios retrieved from observations refer to the same region. They are reported as diamonds in the bottom left panel.}

For $K_{zz}\le 10^{6}$ cm$^{2}$s$^{-1}$, the impact on the composition is minimal \textcolor{black}{with respect to CE. Higher $K_{zz}$ produce instead a clear signature in the VMR of the tracked gases. For four of them (H$_2$O, CH$_4$, CO$_2$ and H$_2$S), increasing the mixing produces a better match with respect to observations. In particular, moving from CE to $K_{zz}=10^{10}$ cm$^{2}$s$^{-1}$ decreases the spread between the retrieved and the modeled VMR from 0.7 to 0.1 dex for H$_2$O, from 3.4 to 2.0 for CH$_4$, from 1.7 to 0.9 for CO$_2$ and from 0.35 to 0.14 for H$_2$S. The case of HCN is more nuanced: the equilibrium VMR is 2.4 dex below the observed one, but at the maximum mixing intensity is 1 dex above. A better match is achieved when $10^8 < K_{zz}  < 10^9$ cm$^{2}$s$^{-1}$. On the other hand, there are two gases whose VMR is better explained by an unmixed atmosphere: CO and NH$_3$. For the former, the spread with the observations increases from 0.05 to 0.35 dex. For the latter, it increases from 4 to 4.9 dex.}

\textcolor{black}{An even stronger $K_{zz}$ would move the modeled VMR of H$_2$O, CH$_4$, CO$_2$ and H$_2$S further toward the retrieved value, while not making those of other gases radically worse. Advecting up into the observable region the gas concentrations at $P=30$ bar would make the spread between the model and the observations for H$_2$O and CO$_2$, respectively equal to 0.03 and 0.09 dex. However, CH$_4$ would still be around 0.85 dex more abundant. This might signal a steeper temperature gradient with an hotter interior, coupled with a very strong mixing. Alternatively, it may indicate that the retrieved carbon abundance is higher than the actual one, since all the C-bearing species tracked here are overabundant with respect to the observed ones.}

As far as the Si and Mg partitions are concerned, we find \textcolor{black}{them mainly in the form of SiO and atomic Mg, respectively, even though a substantial fraction is still in} enstatite and forsterite at the retrieved cloud top level \citep[$2\times 10^{-2}$ bars,][]{hoch25}. \textcolor{black}{SiH$_4$ increases by $\sim$16 orders of magnitude with respect to CE, up to a VMR of $\sim$$3\times10^{-10}$ for $K_{zz}=10^{10}$ cm$^{2}$s$^{-1}$. However, an even stronger mixing pushing the quenching point deeper would enhance its concentration to $\gtrsim$$3\times10^{-8}$, making it clearly detectable \citep[][]{faherty25}.} \textcolor{black}{Focusing on the silicates, we notice that} the fractional abundance of MgSiO$_3$ in the mix changes as a function of $K_{zz}$ and the specified atmospheric layer \textcolor{black}{by enriching the cloud deck of forsterite}, as shown in Figure \ref{fig:yses1c}, bottom right panel, \textcolor{black}{and that for $K_{zz} < 10^7$ cm$^2$s$^{-1}$ the uppermost layers probed are depleted of suspended particles. This depletion is in line with the estimates in Sect.~\ref{ss:settling}, since the 1 mbar layer is the pressure at which the settling and the eddy mixing velocities becomes equal for 0.1 $\mu$m grains and $K_{zz}=10^8$ cm$^2$s$^{-1}$. As far as the composition is concerned, the case with $K_{zz}= 10^{10}$ cm$^2$s$^{-1}$ does not match the observationally retrieved composition of the cloud deck. These facts, taken together, constrain the eddy diffusion in the $10^8 \le K_{zz} \le 10^9$ cm$^2$s$^{-1}$ range.}

%%%%%%%%%%%%%%%%%%%%%%%%%%%%%%%%%%%%%%%%%%%%%%%%%%%%%%%%%%%%%%%%%%
%%%%%%%%%%%%%%%%%%%%%%%%%%%%%%%%%%%%%%%%%%%%%%%%%%%%%%%%%%%%%%%%%%
\section{Discussion}
\label{sec:discussion}

\subsection{Comparison with other CE models}
\label{ss:comparison_ce}

\textcolor{black}{\texttt{SILANE} produces consistent results when compared to a CE model tracking the same set of species, as shown in Sect.~\ref{sec:comparison}. However, it is worth checking how it fares against models that include a larger number of condensates.} 

\textcolor{black}{As far as the Si partition is concerned, the results for both the $10^{-4}$ and the 1 bar atmospheric layers (upper and lower panels of Fig.~\ref{fig:Si_partition}) are in good qualitative agreement with previously published simulations. In particular, the three regimes identified in Sect.~\ref{ss:Si_partition} and the transition temperatures at the $10^{-4}$ bar level are within less than 100 K from those found by \cite{fegley10} and shown in their Fig.~7. Similarly, the 1 bar level results (lower left panel of Fig.~\ref{fig:Si_partition}) are comparable with those shown in Fig.~8 of \cite{visscher10}, provided that the difference between the $T_{eq}$ and the actual temperature of the 1-bar layer is taken into account. On the other hand, changing the set of included minerals inevitably leads to a different partition of Si among its carriers. \cite{fegley10} tracked condensates such as NaAlSiO$_3$O$_8$ (albite), MgCaSi$_2$O$_6$ (diopside) and CaAlSi$_2$O$_8$ (anorthite) and found that they are expected to capture between 30\% ($T_{eq} \le 900$ K) and 12\% ($1000\, \text{K} \le T_{eq} \le 1400$ K) of all the Si atoms at the $10^{-4}$ bar level in a Solar-composition atmosphere. \cite{visscher10} found the same partition at the 1 bar level.}

\textcolor{black}{These differences can be found also in the partition of O, albeit in a more contained fashion. In this case, \cite{fegley10} finds that around 6\% ($T_{eq} \le 900$ K) to 3\% ($T_{eq} \ge 1000$ K) of oxygen is locked in species that we do not track (again, the Ca,Al-silicates). }

As for Fe, \cite{fegley10} predicts that (i) a few \% of it should still be locked in troilite (FeS) at 700 K, even though it then disappears completely at a few tens K higher temperature and that (ii) around 4\% should be in Fe$_3$P if $T_{eq} < 1300$ K, but again we cannot trace them. Overall however, the impact of missing species on iron is smaller than that on Si in this temperature range and, more importantly it does not impact the partition of O.

\textcolor{black}{Differences between our results} and those from \cite{fonte23} arise \textcolor{black}{instead} from the latter work adopting a CE model without condensates, which implies that most Mg, Si and Fe were available to form gas-phase molecules with volatile elements. In particular, they obtain that most Mg and Fe form Mg(OH)$_2$ and Fe(OH)$_2$, while Si forms SiO. The formation of Fe(OH)$_2$ is a major deviation from our model, since when iron is allowed to condense it essentially never forms molecules with other species in the [700, 1600] K temperature interval. On the whole, these three species are expected to capture up to 26\% of oxygen at $\sim 1$ Z$_\odot$ and $T_{eq} = 700$ K, and around 6\% of the oxygen at $T_{eq} \ge 1100$ K, where only SiO remains undissociated. While a perfect comparison cannot be drawn since \cite{fonte23} did not adopted a Solar composition, both the sequestered oxygen fraction and the transition temperature at which refractory carriers disappear are similar, even if the carriers are different.

\subsection{Comparison with other CK models}
\label{ss:comparison_ck}

A more consistent comparison can be drawn with CK models of hot Jupiter atmospheres that include Si carriers. Due to their relevance with respect to our investigation, we will discuss the results of the gas-phase only model of \cite{agundez25} and the gas+condensate model of \cite{kiefer24}.

\cite{agundez25} investigated the disequilibrium chemistry and its feedback on the vertical temperature profile in the atmospheres of HD 209458 b and HD 189733 b, and includes several Si species including the Si hydrides, SiO, SiO$_2$, SiS and SiC, but no Si condensates. Their radiative-chemistry model has been run on the dayside of these planets using a Solar elemental composition and considering a vertically variable $K_{zz}$ in the 10$^{9}$-$10^{11}$ cm$^2$ s$^{-1}$ range. They find that in both HD 209458 b and HD 189733 b Si is expected to be mostly locked in SiO, which is in accord with our results since the temperature profiles for both planets are hotter than our profile in the $T_{eq}=1300$ K case. However, their S is nearly entirely locked in H$_2$S and thus it is not available to form SiS, as we find in this study. This difference is possibly due to a combination of a different temperature profile, a one-to-two orders of magnitude smaller vertical mixing, and a larger set of reactions involving SiS in their model. While the \cite{agundez25} Si network is larger than ours, it does include several reactions whose Arrhenius coefficients have been derived at room temperature, i.e. substantially outside of the range typical for hot Jupiter atmospheres.

\cite{kiefer24} studied on the formation of mineral clouds in exoplanetary atmospheres. Their model involves detailed nucleation and grain growth models that include both condensation and surface reactions, allowing for the formation, among other silicates and oxides, of forsterite. They focused in particular on the timescale of formation of the TiO$_2$ cloud condensation nuclei and the subsequent evolution of the grain composition at different longitudes (morning and evening terminator, substellar and antistellar points) and depths (between 0.002 and 2 bars) on HD 209458 b. However, they did not include the impact of vertical mixing, since they run their model at fixed positions along the atmospheric pressure-temperature profiles. They found that virtually all Mg is locked in forsterite at the 0.002 bar level for $T\lesssim1400$ K, as it is a majority of Si. Forsterite is also the main Si and Mg carrier at the 2 bar level for $T\sim1800$ K. As far as Fe is concerned, \cite{kiefer24} finds it locked mostly in fayalite (Fe$_2$SiO$_4$), or Fe$_2$O$_3$ for $T\gtrsim 1400$ K. These results differ substantially both with respect to ours and those of previous literature, especially in the case of iron, which is generally expected to remain in elemental form, both in CK and in CE models. This helps identify the source of the differences between their results and ours in the chemical network, and in particular in the inclusion of surface reactions. In fact, while metallic iron is the thermodynamically favored form of this element in the considered temperature range, its condensation requires a strong supersaturation and is slower than the formation of other Fe carriers \citep[see e.g.][]{tielens22}. Similarly, forsterite can condense at several hundreds of K higher \citep[e.g.][]{hu24} than that found by pure thermochemical considerations once an appropriate substrate is present.

\subsection{\textcolor{black}{Limitations of the grain treatment}}
\label{ss:limitations}

%\textcolor{black}{This work has two important limitations that should be kept in mind when considering our results. First, CCNs are not built self-consistently in the code but always present, and condensation is treated as a simple phase change from virtual silicate gases. Second, grain size is limited to a single, constant value for each species. We chose a common $r_p=1$ $\mu$m for all three condensates. Actual grain sizes likely vary by 4-5 orders of magnitude along the column, potentially reaching $\sim$100 $\mu$m at $P>100$ bar \citep[][]{woitke03,arfaux24}. Larger grains have higher settling velocities, which likely lead to the accumulation of condensates at depths higher than the maximum value explored here (1000 bar), and produce a stronger depletion of condensates in the upper parts of the atmosphere than the one reported here. However, including these phenomena will require a substantial expansion of VULCAN, and will be the focus of a future work on this subject.}

\textcolor{black}{This work has two important limitations that should be kept in mind when considering our results. First, CCNs are not built kinematically in the code but always present, and condensation is treated as a simple phase change from virtual silicate gases. Second, grain size is limited to a single, constant value for each species. We chose a common $r_p=1$ $\mu$m for all three condensates for our idealized Jupiter-like planet, and $r_p=0.1$ $\mu$m to model the real exoplanet YSES-1 c. However, actual grain sizes likely vary by 4-5 orders of magnitude along the column, potentially reaching $\sim$100 $\mu$m at $P>100$ bar \citep[][]{woitke03,arfaux24}. Larger grains have higher settling velocities, which lead to the accumulation of condensates at depths higher than the values explored here, and produce a stronger depletion of condensates in the upper parts of the atmosphere than the one reported in Section \ref{sec:results}. However, including these phenomena will require a substantial expansion of VULCAN, and will be the focus of a future work on this subject.}

\section{Conclusions}
\label{sec:conclusions}

In this work, we present a new chemical network, called \texttt{SILANE}, for the CK code VULCAN that includes Mg, Si, Fe, and 15 of their most important carriers expected to from in hot Jupiter atmospheres, including the two rock-forming silicates MgSiO$_3$ and Mg$_2$SiO$_4$. This network has been constructed using the data on high temperature gas-phase reactions involving refractories in the NIST repository, in \cite{decin18} \textcolor{black}{\cite{zanchet18} and \cite{rosi18}}, plus ad hoc prescriptions for the dissociation of MgSiO$_3$ and Mg$_2$SiO$_4$. \textcolor{black}{To be able to include condensed silicates in VULCAN, we had to consider a virtual gas-phase for both enstatite and forsterite, and to disable the reverse of the enstatite- and forsterite-forming reactions.} We tested \texttt{SILANE} in the [700, 1600] K equilibrium temperature range by comparing the resulting abundances with those derived by a CE model, FastChem3 by disabling all the sources of chemical disequilibrium (Sect.~\ref{sec:comparison}). With respect to the CE model, the mixing ratios produced by \texttt{SILANE} \textcolor{black}{show only small differences in major carriers. As such, we consider it to be suitable to investigate the partition of these elements at the current level of observational accuracy.}

We then applied \texttt{SILANE} to the study of a vertically mixed, Solar-composition exoplanetary atmosphere. The eddy diffusion coefficient has been set to $10^{8}$ cm$^2$ s$^{-1}$ and sedimentation via gravitational settling was enabled. We find that
\begin{itemize}
\item enstatite and forsterite disappear at a far lower $T_{eq}$ (1000 K) than expected from CE calculations (1400 K and 1500 K), \textcolor{black}{and are substituted mainly by SiO, \textcolor{black}{Mg(OH)$_2$} and Mg};
\item SiH$_4$ VMR is boosted by up to $\sim$20 orders of magnitude in a vertically mixed atmosphere and can reach ppm-level concentrations in the 800-1100 K range, up to a maximum of \textcolor{black}{$\sim10^{-7}$ for $T_{eq} = 1000$ K}, making it a marker of a strong vertical mixing;
\item the amount of O locked in refractories is $\sim$21\% at $T_{eq} = 700$ K and reaches $\sim$6\% at $T_{eq} = 1000$ K, due to the dissociation of the rock-forming silicates. This behavior is substantially different from that predicted by CE models, which predict either an higher fraction of oxygen as locked in refractories \citep[$\sim$26\% at $Z_\odot \simeq 1$,][]{fonte23}, or a persistence of the same sequestered oxygen fraction up to higher temperatures \citep[$\sim$1400 K,][]{fegley10,visscher10}. The comparison with \cite{fonte23} is not exact however, since the composition used in that work deviates from the Solar one.
\end{itemize}
%These results have been applied to the recent observations of enstatite-rich silicate clouds in YSES-1 c \citep{hoch25} to constrain the vertical mixing of its atmosphere. Since enstatite is present at the $10^{-4}$ bar level in chemical equilibrium but disappears in mixed atmospheres already at $\sim 900$ K, its detection points to a non-mixed or weakly mixed atmosphere on the vertical direction.

Finally, we applied \texttt{SILANE} to the study of the young hot Jupiter YSES-1 c, whose recent observations allowed both to retrieve the abundances of several volatile species and to prove the presence of enstatite-forsterite clouds \citep[][]{hoch25}. \textcolor{black}{Our results in terms of gas phase VMRs} point to a well-mixed atmosphere, possibly with \textcolor{black}{$k_{zz} \gtrsim 10^{10}$ cm$^2$s$^{-1}$. In particular, this level of $K_{zz}$ reproduces well the observed VMR of H$_2$O and CO, and partially closes the gap between the observed and expected VMRs of CH$_4$ and CO$_2$. On the other hand, our results in terms of condensate budget and composition favors a $K_{zz}$ in the [10$^{8}$,10$^{9}$] cm$^2$s$^{-1}$ range, constrained at the lower limit by the need of a sufficient diffusion to keep grains aloft, and at the upper limit by the observationally} retrieved mixture of enstatite and forsterite in the [0.01, 0.03] bar region.
%which is also in better agreement with the retrieved mixture of enstatite and forsterite in the [0.01, 0.03] bar region, i.e. that of the cloud base.

\begin{acknowledgements}
P.S. and D.P. are supported by the National Research Centre for High Performance Computing, Big Data e Quantum Computing CN\_00000013, Spoke 3 `Astrophysics and Cosmos Observations' (CUP: C53C22000350006). The authors acknowledge the support of the Italian National Institute of Astrophysics (INAF) through the Main Stream project `Ariel and the astrochemical link between circumstellar discs and planets' (CUP: C54I9000700005) and the PRIN INAF 2019 `Platea' and `Hot-Atmos', of the Italian Space Agency (ASI) through the ASI-INAF contracts N.~2016-23-H.0 plus addendum 2016-23-H.2-2021 and 2021-5-HH.0 plus addenda 2021-5-HH.1-2022 and 2021-5-HH.2-2024 and of the European Research Council via the Horizon 2020 Framework Programme ERC Synergy `ECOGAL' Project GA-855130. The authors also acknowledge the support of the European Cooperation in Science and Technology (COST) Action CA22133 'The birth of solar systems (PLANETS)'.
\end{acknowledgements}

%-------------------------------------------------------------------

\bibliographystyle{mnras}
\bibliography{biblio}

\onecolumn
\begin{appendix}
\section{\textcolor{black}{Selection of species to include in the network}}

\begin{figure*}
\centering
%\isPreprints{\centering}{} % Only used for preprints
\includegraphics[width=0.33\textwidth]{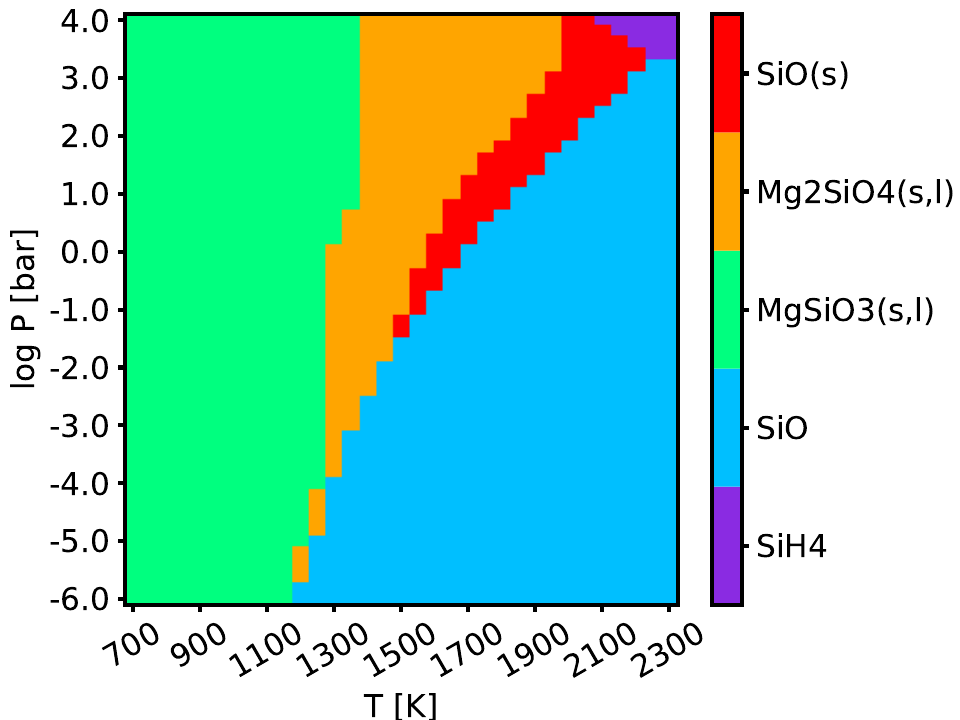}
\includegraphics[width=0.33\textwidth]{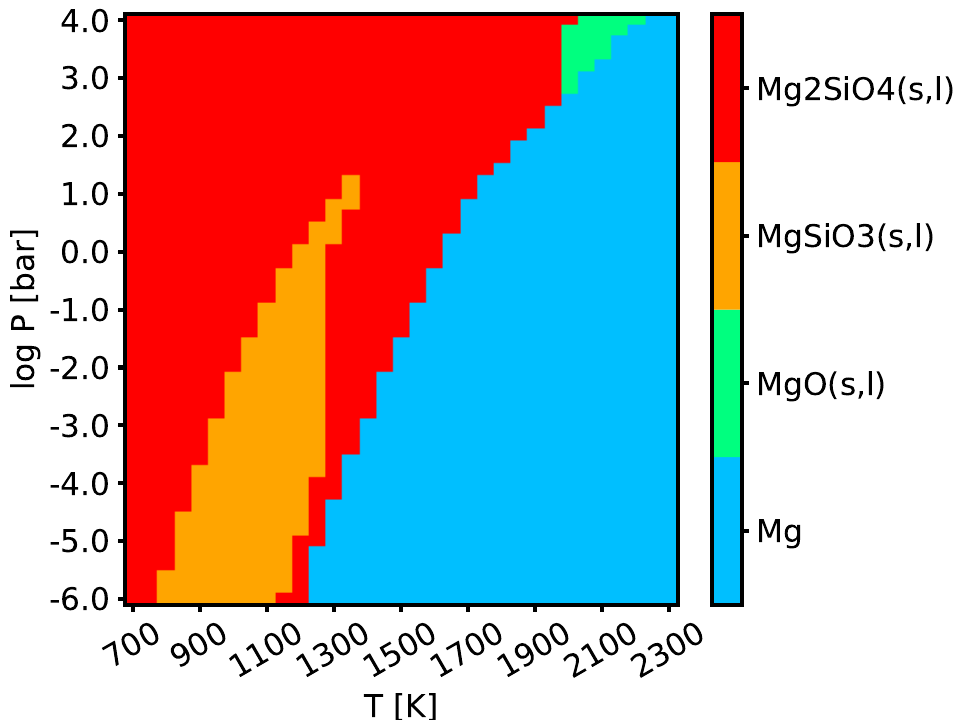}
\includegraphics[width=0.33\textwidth]{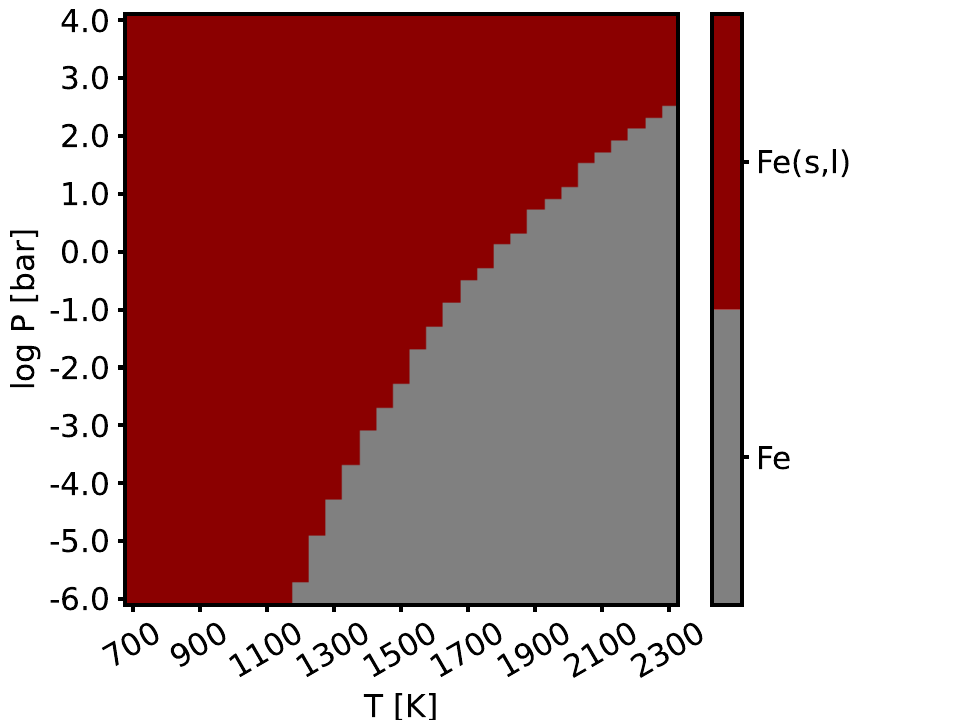}\\
\includegraphics[width=0.33\textwidth]{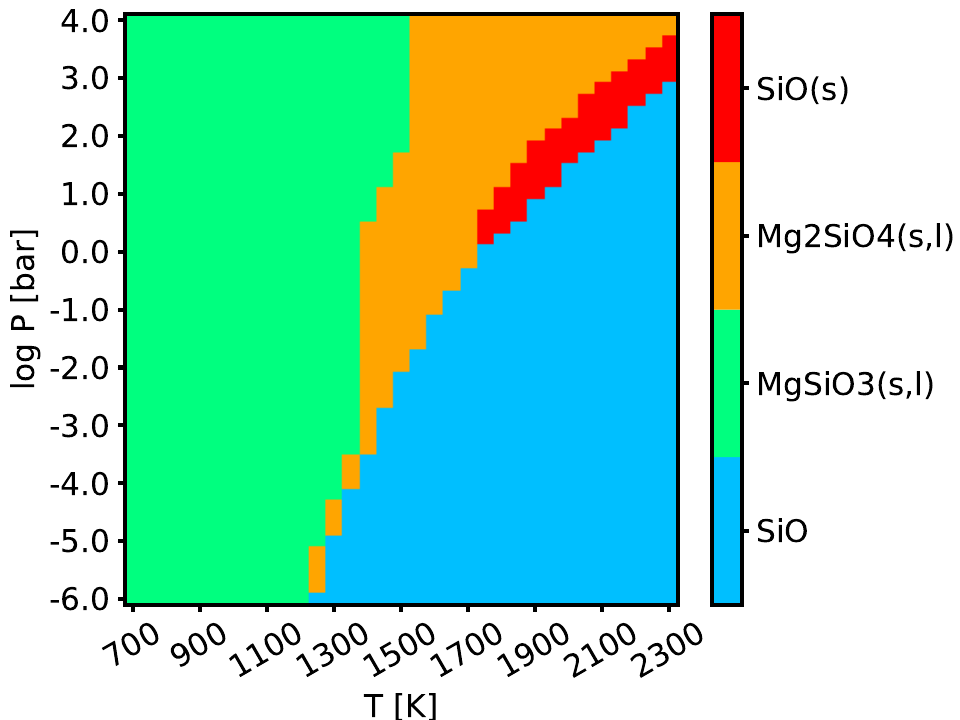}
\includegraphics[width=0.33\textwidth]{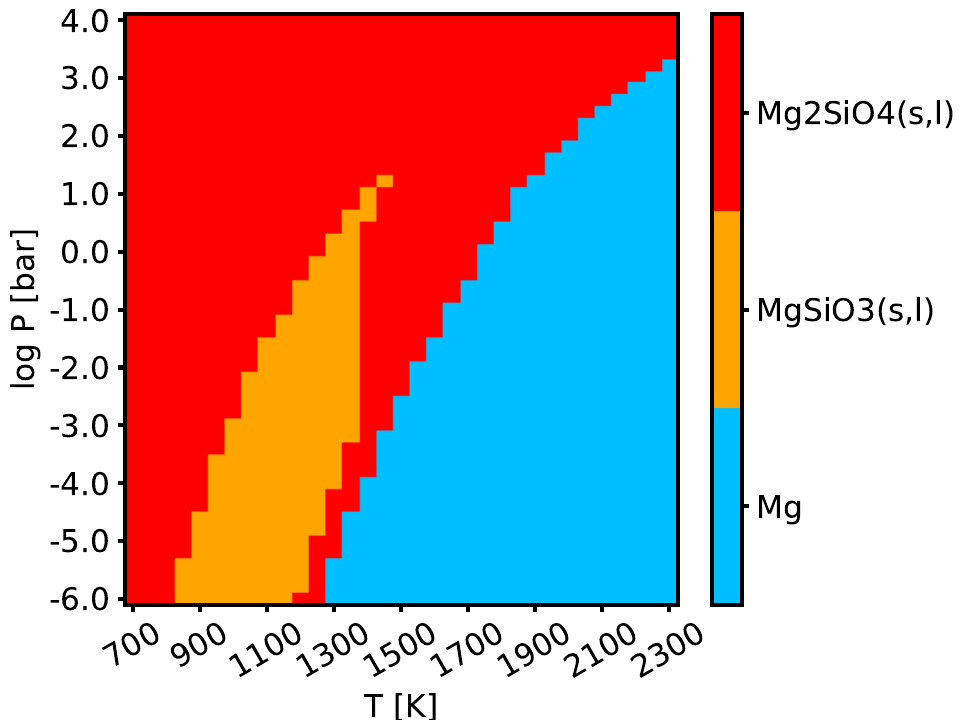}
\includegraphics[width=0.33\textwidth]{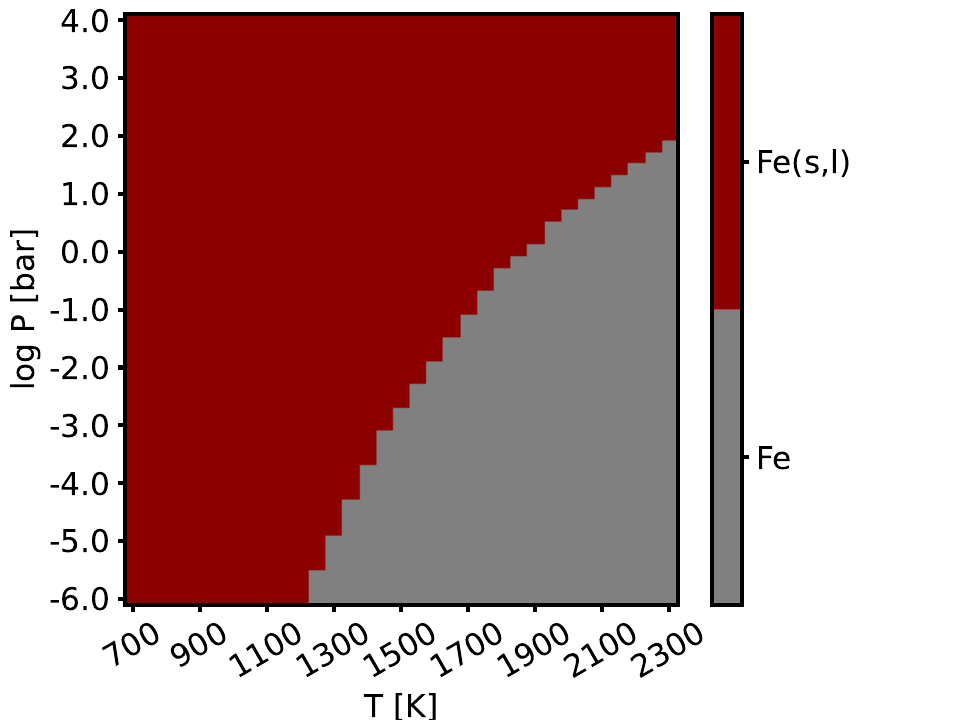}\\
\includegraphics[width=0.33\textwidth]{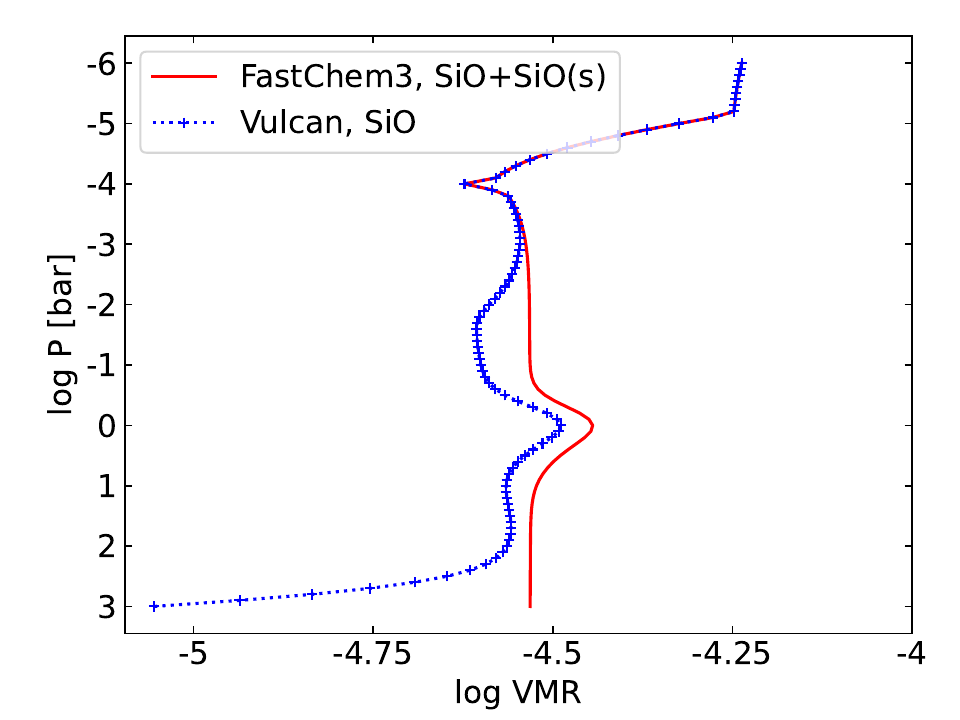}
\includegraphics[width=0.33\textwidth]{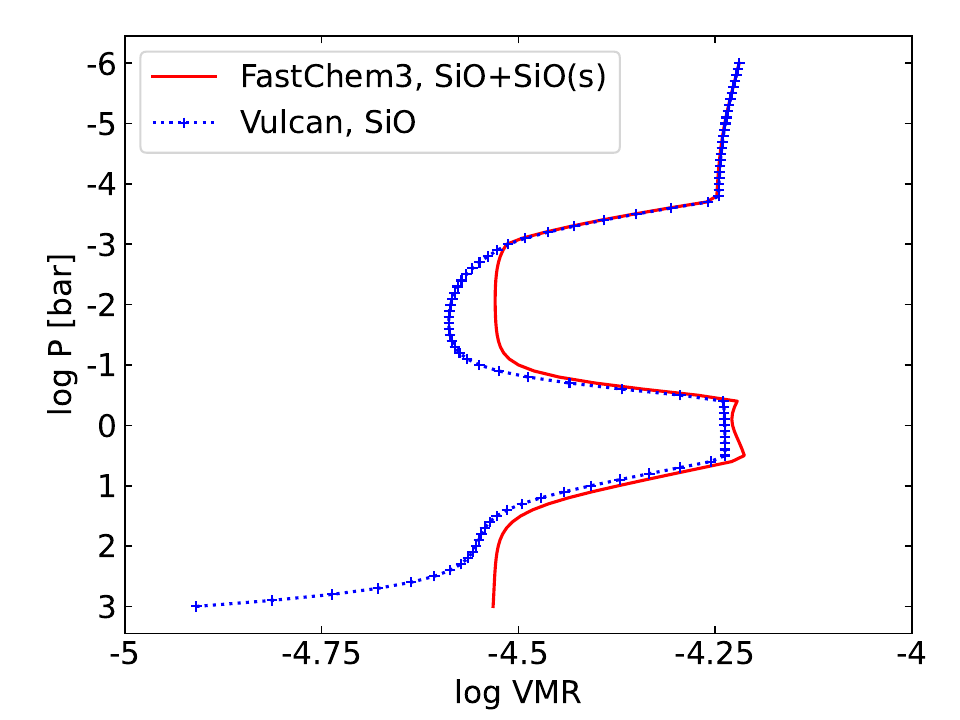}
\includegraphics[width=0.33\textwidth]{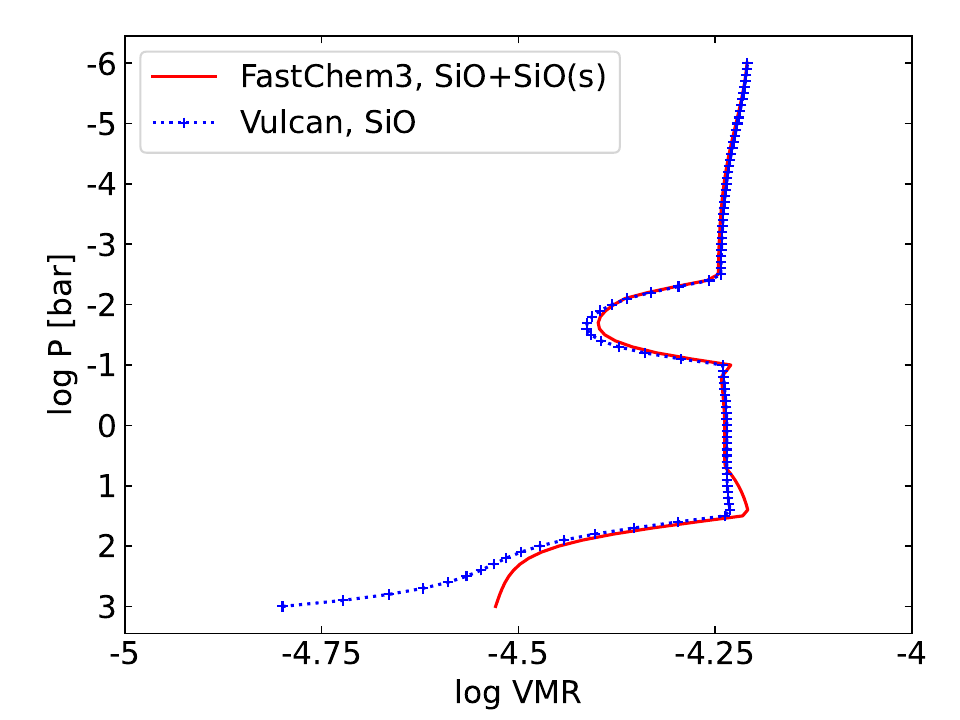}\\
\caption{\textcolor{black}{Results of the selection process to identify the condensates to include in \texttt{SILANE}.} Upper row, from left to right: the dominant Si, Mg and Fe carriers at chemical equilibrium as calculated by FastChem3 \citep[][]{kitzmann24} for a Solar composition gas. Centre: same as for the upper row, but for a $3.3\times$ Solar composition. Lower row, form left to right: the comparison between the vertical distributions of gaseous and solid SiO as calculated by FastChem3 in red, and gaseous SiO as calculated by VULCAN with all sources of chemical disequilibrium turned off in blue, for $T_{eq}=1400$ K, 1500 K and 1600 K. \label{fig:dominant_carriers}}
\end{figure*}

\textcolor{black}{In order to confirm that the set of species included in the network is actually representative of the main carriers of Mg, Si, and Fe, we run FastChem3 on a grid of pressures and temperatures relevant to the cases we want to study with our new network and identified, for each (p,T) point, the main gas or condensate carriers. The results are shown in Fig.~\ref{fig:dominant_carriers}, upper (Solar metallicity) and centre ($3.3\times$ Solar metallicity) rows. As it is possible to see, there are only two species that are expected to form abundantly but are not present in the network: the condensed phases of MgO and SiO, hereafter called MgO(s,l) and SiO(s). The condensate form of SiO$_2$, which has been observed \citep[][]{grant23}, never forms in the portion of the pressure-temperature-metallicity parameter space that we tested here.}

\textcolor{black}{As far as MgO(s,l) is concerned, its formation happens at $P\gtrsim300$ bar and $T\gtrsim2000$ K, which is a region of the parameter space that we do not explore in this paper, since the bottom of the atmospheric columns generated for the hottest case examined here, i.e. a Solar-composition gas giant with $T_{eq}=1600$ K, is at 1860 K (see Fig.~\ref{fig:pt_profiles}, dark red curve).}

\textcolor{black}{As fas as SiO(s) is concerned, we initially calculate its abundance in chemical equilibrium, but then we treat all SiO (both gaseous and solid) as gaseous for the purpose of chemical kinetics. In practice, the abundances of SiO and SiO(s) are summed together just before being handed over to VULCAN. In principle, the chemical behavior of SiO and SiO(s) are different. As such, we check if the vertical mixing ratio profiles of SiO+SiO(s) (as obtained by FastChem3), and SiO (as obtained by VULCAN by turning off all sources of chemical disequilibrium) are similar. The results of this comparison are shown in Fig.~\ref{fig:dominant_carriers}, lower panels, for $T_{eq}= 1400$ K, 1500 K and 1600 K, i.e. in the cases the VMR of SiO is sufficiently high in chemical equilibrium to allow for the formation of SiO(s). The mixing ratios obtained with this methodology are very similar for $P<100$ bar, where the maximum difference is $\sim$12\%, and diverge only in the bottom part of the atmosphere, where differences up to $\sim$0.5 dex may arise. These differences are most likely related to the lack the inverse reactions 1117 and 1119, which means that the network cannot perfectly match equilibrium chemistry results. We also remark that the rates of reactions involving SiO (907, 909, 911, 913, 915 and 917 in Table \ref{tab:bimolecular}) are all derived in an higher temperature interval than that explored in this work, but one, although this would not necessarily lead to discrepancies between CE and CK values. We, however, underline how these differences have seemingly a minimal impact on the partition of the observable O among its carriers, as discussed in Sect.~\ref{ss:O_partition}. Since the main scientific goal of this network is to capture the sequestration of oxygen by refractories to enhance the estimates of the C/O ratio, we consider this difference acceptable.}

\section{Reactions included in the chemical network}

Reaction rates are taken from the United States National Institute for Standards and Technology (NIST), from \cite{decin18}, \textcolor{black}{\cite{zanchet18} and \cite{rosi18},} or set ad-hoc because of the lack of published data. The ID of the reaction refers to the one assigned internally in our chemical network and it is used to cite the reaction in this paper wherever necessary. References are reported as in the NIST database to facilitate the comparison with the published data.

\begin{table*}
\centering
\begin{threeparttable}
\small % Change table font size
%\begin{adjustwidth}{-\extralength}{0cm}
%\caption{The 37 second-order reactions involving the refractories Mg, Si and Fe and the volatiles H, C, N and O considered in this paper. $\mathcal{A}$ is in cm$^3$ molecule s$^{-1}$, $T_a = E_a/R$ (where $E_a$ is the activation energy and $R$ is the ideal gas constant), $T_{min}$ and $T_{max}$ are in K. Notice that \cite{decin18} does not provide the temperature interval within which the coefficients were derived. Silicate-forming reactions 1117 and 1119 are never reversed. \label{tab:bimolecular}}
\caption{Second-order reactions involving Mg, Si and Fe. \label{tab:bimolecular}}
%\begin{tabularx}{\fulllength}{llcccll}
\begin{tabularx}{\linewidth}{llcccll}
\toprule
ID & Reaction & $\mathcal{A}$ & $b$ & $T_a$ & $T_{min}-T_{max}$ & Reference \\
\midrule
875 & Fe + H$_2$O $\rightarrow$ FeO + H$_2$   & $8.80\times10^{-11}$ & 0.0   & 11925.0 & n.a. & \cite{decin18}, R40  \\
877 & Fe + OH $\rightarrow$ FeO + H           & $2.40\times10^{-11}$ & 0.0   & 3348.0  & n.a. & \cite{decin18}, R42 \\
879 & Fe + N$_2$O $\rightarrow$ FeO + N$_2$   & $1.99\times10^{-10}$ & 0.0   & 5493.4  & 870-1820  & 1998ZAS/SMI449-450\\
881 & Fe + O$_2$ $\rightarrow$ FeO + O        & $1.79\times10^{-9}$ & -0.02  & 11229.8 & 200-2500  & 2011SMI166-169 \\
883 & Fe + CO$_2$ $\rightarrow$ FeO + CO      & $1.86\times10^{-17}$ & 1.98  & 8349.0  & 1000-3000 & 2020VIC/DEL16943-16948 \\
885 & FeO + NO$_2$ $\rightarrow$ FeO$_2$ + NO & $5.04\times10^{-10}$ & 0.0   & 438.0   & 196-527   & 1999PLA/ROL1843-1849 \\
887 & FeO + O$_2$ $\rightarrow$ FeO$_2$ + O   & $1.04\times10^{-12}$ & 0.40  & 8204.6  & 1000-3000 & 2012SMI543-553 \\
889 & FeO + CO$_2$ $\rightarrow$ FeO$_2$ + CO & $6.48\times10^{-9}$  & 0.0   & 22984.0 & 1330-2650 & 2002GIE/HER3665-3668 \\
\midrule
891 & Mg + H$_2$O $\rightarrow$ MgO + H$_2$   & $5.80\times10^{-10}$ & 0.0 & 29435.0 & n.a. & \cite{decin18}, R48 \\
893 & Mg + H$_2$O $\rightarrow$ MgOH + H      & $1.10\times10^{-8}$  & 0.0 & 23173.0 & n.a. & \cite{decin18}, R50 \\
895 & Mg + OH $\rightarrow$ MgO + H           & $3.40\times10^{-10}$ & 0.0 & 23651.0 & n.a. & \cite{decin18}, R52 \\
897 & Mg + CO$_2$ $\rightarrow$ MgO + CO      & $3.00\times10^{-9}$  & 0.0 & 33054.0 & n.a. & \cite{decin18}, R54 \\
899 & Mg + O$_2$ $\rightarrow$ MgO + O        & $3.00\times10^{-9}$  & 0.0 & 26584.0 & n.a. & \cite{decin18}, R56 \\
901 & Mg + NO$_2$ $\rightarrow$ MgO + NO      & $5.50\times10^{-12}$ & 0.0 & 409.1   & 873-1020 & 2001ROL/PLA733-4740 \\
903 & Mg + N$_2$O $\rightarrow$ MgO + N$_2$   & $2.16\times10^{-11}$ & 0.0 & 3321.3  & 823-1070 & 2001ROL/PLA733-4740 \\
\midrule
905 & Si + SH $\rightarrow$ SiS + H           & $5.87\times10^{-9}$  & -0.76 & 9.6    & 25-1000   & 2021MOT/VAR37 \\
907 & Si + NO $\rightarrow$ SiO + N           & $5.31\times10^{-11}$ & 0.0   & 1780.9 & 1660-3660 & 1993MIC/MAT6839-6842 \\
909 & Si + N$_2$O $\rightarrow$ SiO + N$_2$   & $1.33\times10^{-10}$ & 0.0   & 0.0    & 1780-3560 & 1994MIC/ROT5310-5313 \\
911 & Si + OH $\rightarrow$ SiO + H           & $3.32\times10^{-10}$ & 0.0   & 0.0    & 1780-3560 & 1994MIC/ROT5310-5313 \\
913 & Si + CO $\rightarrow$ SiO + C           & $1.30\times10^{-9}$  & 0.0   & 34536.7& 1780-3560 & 1994MIC/ROT7844-7847 \\
915 & Si + CO$_2$ $\rightarrow$ SiO + CO      & $9.96\times10^{-10}$ & 0.0   & 9424.8 & 2100-3190 & 1994MIC/ROT7844-7847 \\
\textcolor{black}{917} & Si + SO $\rightarrow$ SiS + O           & $1.77\times10^{-11}$ & 0.16  & -20.0   & 100-1500  & \cite{zanchet18} \\
\textcolor{black}{919} & Si + SO $\rightarrow$ SiO + S           & $1.53\times10^{-10}$ & -0.11 & 32.0    & 100-2500  & \cite{zanchet18} \\
\textcolor{black}{921} & SiS + O $\rightarrow$ SiO + S           & $9.53\times10^{-11}$ & 0.29 & -32.0    & 50-2500   & \cite{zanchet18} \\
923 & SiO + OH $\rightarrow$ SiO$_2$ + H      & $2.98\times10^{-14}$ & 0.78  & 613.7  & 1000-2500 & 1994MIC/ROT7844-7847 \\
\textcolor{black}{925} & SiH + S $\rightarrow$ SiS + H           & $1.00\times10^{-10}$ &  0.00 & 0.0     & n.a.     & \cite{rosi18}, assumed  \\
927 & SiH + H $\rightarrow$ Si + H$_2$            & $1.20\times10^{-15}$ & 1.65  & -4.8   & 300-2000  & 2013RAG/LEE1735-1746 \\
929 & SiH$_2$ + H $\rightarrow$ SiH + H$_2$       & $1.05\times10^{-15}$ & 1.82  & 99.9   & 300-2000  & 2013RAG/LEE1735-1746 \\
931 & SiH$_3$ + H $\rightarrow$ SiH$_2$ + H$_2$   & $1.86\times10^{-08}$ & -0.54 & 593.3  & 1000-2000 & 2013RAG/LEE1735-1746 \\
933 & SiH$_3$ + H$_2$ $\rightarrow$ SiH$_4$ + H   & $2.49\times10^{-16}$ & 1.90  & 1101.1 & 290-2000  & 2013RAG/LEE1735-1746 \\
935 & SiH$_4$ + O $\rightarrow$ SiH$_3$ + OH      & $1.53\times10^{-17}$ & 2.25  & 795.4  & 200-3000  & 2013RAG/LEE1735-1746 \\
937 & SiH$_4$ + OH $\rightarrow$ SiH$_3$ + H$_2$O & $4.04\times10^{-18}$ & 2.55  & -142.0 & 200-3000  & 2013RAG/LEE1735-1746 \\
939 & SiH$_4$ + CH$_3$ $\rightarrow$ SiH$_3$ + CH$_4$& $1.43\times10^{-24}$ & 3.95  & 2876.1 & 600-2000  & 2013RAG/LEE1735-1746 \\
941 & SiH$_4$ + Si $\rightarrow$ SiH$_2$ + SiH$_2$& $1.71\times10^{-20}$ & 2.76  & 5932.6 & 300-1600  & 2013RAG/LEE1735-1746 \\
943 & SiH$_4$ + Si $\rightarrow$ SiH$_3$ + SiH    & $1.03\times10^{-15}$ & 1.33  & 11054.2& 300-1600  & 2013RAG/LEE1735-1746 \\
\midrule
1117 & MgSiO$_3$ + M $\rightarrow$ MgO + SiO$_2$ + M                & $1.00\times10^{-10}$  & 0.0 & 0.0 & n.a. & this work \\
1119 & Mg$_2$SiO$_4$ + M $\rightarrow$ MgSiO$_3$ + MgO + M          & $1.00\times10^{-10}$  & 0.0 & 0.0 & n.a. & this work \\
\bottomrule
\end{tabularx}
%\noindent{\footnotesize{\textsuperscript{1}1998ZAS/SMI449-450, \textsuperscript{2}2011SMI166-169, \textsuperscript{3}2020VIC/DEL16943-16948, \textsuperscript{4}1999PLA/ROL1843-1849, \textsuperscript{5}2012SMI543-553, \textsuperscript{6}2002GIE/HER3665-3668, \textsuperscript{7}2001ROL/PLA733-4740, \textsuperscript{8}2021MOT/VAR37, \textsuperscript{9}1993MIC/MAT6839-6842, \textsuperscript{10}1994MIC/ROT5310-5313, \textsuperscript{11}1994MIC/ROT7844-7847, \textsuperscript{12}2013RAG/LEE1735-1746.}}
%\end{adjustwidth}
\begin{tablenotes}[flushleft]
\small
\item $\mathcal{A}$ is in cm$^3$ molecule s$^{-1}$, $T_a = E_a/R$ (where $E_a$ is the activation energy and $R$ is the ideal gas constant), $T_{min}$ and $T_{max}$ are in K. Notice that \cite{decin18} do not provide the temperature interval within which the coefficients were derived. Silicate-forming reactions 1117 and 1119 are never reversed.
\end{tablenotes}
\end{threeparttable}
\end{table*}

\begin{table*}
\centering
\begin{threeparttable}
\small % Change table font size
%\begin{adjustwidth}{-\extralength}{0cm}
%\caption{The 5 dissociation reactions involving the refractories Mg, Si and Fe and the volatiles H, C, N and O considered in this paper. The second row, where present, is the limiting reaction rate at infinite pressure. $\mathcal{A}$ is in cm$^3$ molecule s$^{-1}$ in the first row, and s$^{-1}$ in the second row. \label{tab:onemolecular}}
\caption{Dissociation reactions involving Mg, Si and Fe. \label{tab:onemolecular}}
\begin{tabularx}{\linewidth}{llcccll}
\toprule
ID & Reaction & $\mathcal{A}$ & $b$ & $T_a$ & $T_{min}-T_{max}$ & Reference \\
\midrule
1075 & SiH$_2$ + M $\rightarrow$ Si + H$_2$ + M  & $1.64\times10^{-3}$ & -1.76 & 19253.9 & 1200-1700 & 1994MIC/ROT439-451 \\
     &                                           & $9.73\times10^{8}$  & 0.0   & 31407.9 & 300-1000 & 1994JOH/EKE2135-2140 \\
1077 & SiH$_3$ + M $\rightarrow$ SiH$_2$ + H + M & $2.54\times10^{4}$  & -3.14 & 35619.7 & 300-2000 & 2013RAG/LEE1735-1746 \\
     &                                           & $1.28\times10^{15}$ & 0.15 & 35138.4 & 300-2000 & 2013RAG/LEE1735-1746 \\
1079 & SiH$_3$ + M $\rightarrow$ SiH + H$_2$ + M & $7.98\times10^{2}$  & -3.04 & 22262.3 & 300-2000 & 2013RAG/LEE1735-1746 \\
     &                                           & $5.97\times10^{9}$  & 1.23 & 22262.3 & 300-2000 & 2013RAG/LEE1735-1746 \\
1081 & SiH$_4$ + M $\rightarrow$ SiH$_3$ + H + M & $1.23\times10^{12}$  & -4.86 & 47292.4 & 300-2000 & 2013RAG/LEE1735-1746 \\
     &                                           & $1.01\times10^{17}$  & -0.26 & 46209.4 & 300-2000 & 2013RAG/LEE1735-1746 \\
1083 & SiH$_4$ + M $\rightarrow$ SiH$_2$ + H$_2$ + M & $2.41\times10^{9}$  & -4.60 & 29362.2 & 300-2000 & 2013RAG/LEE1735-1746 \\
     &                                               & $2.77\times10^{12}$ & 0.67  & 27677.5 & 300-2000 & 2013RAG/LEE1735-1746 \\
\bottomrule
\end{tabularx}
%\noindent{\footnotesize{\textsuperscript{1}1994MIC/ROT439-451, \textsuperscript{2}1994JOH/EKE2135-2140, \textsuperscript{3} 2013RAG/LEE1735-1746}}
%\end{adjustwidth}
\begin{tablenotes}[flushleft]
\item The second row is the limiting reaction rate at infinite pressure. $\mathcal{A}$ is in cm$^3$ molecule s$^{-1}$ in the first row and in s$^{-1}$ in the second row.
\end{tablenotes}
\end{threeparttable}
\end{table*}

\begin{table*}
\centering
\begin{threeparttable}
\small % Change table font size
%\begin{adjustwidth}{-\extralength}{0cm}
%\caption{The 4 third-order reactions involving the refractories Mg, Si and Fe and the volatiles H, C, N and O considered in this paper. The second row is the limiting reaction rate at infinite pressure. $\mathcal{A}$ is in cm$^6$ molecule$^2$ s$^{-1}$ in the first row, and cm$^3$ molecule s$^{-1}$ in the second row. \label{tab:termolecular}}
\caption{Third-order reactions involving Mg, Si and Fe. \label{tab:termolecular}}
\begin{tabularx}{\linewidth}{llcccll}
\toprule
ID & Reaction & $\mathcal{A}$ & $b$ & $T_a$ & $T_{min}-T_{max}$ & Reference \\
\midrule
1085 & MgO + H$_2$O + M $\rightarrow$ Mg(OH)$_2$ + M & $1.08\times10^{-14}$  & -4.52 & 590.9 & 272-377  & 2001ROL/PLA4733-4740 \\
     &                                               & $3.52\times10^{-10}$  & 0.0   & 334.5 & 272-377  & 2001ROL/PLA4733-4740 \\
1087 & FeO + H$_2$O + M $\rightarrow$ Fe(OH)$_2$ + M & $6.15\times10^{-26}$  & -1.16 & 0.0   & 298-527  & 2000ROL/PLA2335-2343 \\
     &                                               & $5.35\times10^{-10}$  & 0.0   & 611.3 & 298-527  & 2000ROL/PLA2335-2343 \\
1089 & Fe + O$_2$ + M $\rightarrow$ FeO$_2$ + M      & $5.46\times10^{-22}$  & -2.60 & 3170.9& 300-2000 & 1998SMI844-848 \\
     &                                               & $2.59\times10^{-10}$  & 0.0   & 2262.3& 300-2000 & 1998SMI844-848 \\
1091 & SiH + H + M $\rightarrow$ SiH$_2$ + M         & $8.67\times10^{-27}$  & -2.40 & 353.8 & 300-2300 & 2013RAG/LEE1735-1746 \\
     &                                               & $5.71\times10^{-12}$  & 0.36  & 51.7  & 300-2000 & 2013RAG/LEE1735-1746 \\
\bottomrule
\end{tabularx}
\begin{tablenotes}[flushleft]
\item The second row is the limiting reaction rate at infinite pressure. $\mathcal{A}$ is in cm$^6$ molecule$^2$ s$^{-1}$ in the first row and in cm$^3$ molecule s$^{-1}$ in the second row.
\end{tablenotes}
%\noindent{\footnotesize{\textsuperscript{1}2001ROL/PLA4733-4740, \textsuperscript{2}2000ROL/PLA2335-2343, \textsuperscript{3}1998SMI844-848, \textsuperscript{4} 2013RAG/LEE1735-1746}}
%\end{adjustwidth}
\end{threeparttable}
\end{table*}

\end{appendix}

\end{document}